\documentclass[prx,amsmath,amssymb,twocolumn,longbibliography,aps,10pt]{revtex4-2}
\usepackage{graphicx}
\usepackage{subfigure}
\usepackage{adjustbox}
\usepackage{bm}
\usepackage{array}
\usepackage{color}
\usepackage{braket}
\usepackage{standalone}
\usepackage{multirow}
\usepackage{tikz}
\usepackage{mathrsfs}
\usepackage{dsfont}
\usepackage{comment}
\usepackage{amsmath}
\usepackage[colorlinks,bookmarks=true,citecolor=blue,linkcolor=red,urlcolor=blue]{hyperref}
\usepackage[capitalize]{cleveref}
\usepackage[normalem]{ulem}
\usepackage{cancel}

\graphicspath{{./Figs/}}

\newcommand{\mycomment}[1]{}

\begin{document}

\title{
Regularizing 3D conformal field theories via anyons on the fuzzy sphere
}

\author{Cristian Voinea$^{1}$, Ruihua Fan$^{2}$, Nicolas Regnault$^{3,4,5}$ and Zlatko Papi\'c$^{1}$}

\affiliation{${}^{1}$School of Physics and Astronomy, University of Leeds, Leeds LS2 9JT, United Kingdom}
\affiliation{${}^{2}$Department of Physics, University of California, Berkeley, CA 94720, USA}
\affiliation{${}^{3}$Center for Computational Quantum Physics, Flatiron Institute, 162 5th Avenue, New York, NY 10010, USA}
\affiliation{${}^{4}$Department of Physics, Princeton University, Princeton, New Jersey 08544, USA}
\affiliation{${}^{5}$Laboratoire de Physique de l'Ecole normale sup\'{e}rieure, ENS, Universit\'{e} PSL, CNRS, Sorbonne Universit\'{e}, Universit\'{e} Paris-Diderot, Sorbonne Paris Cit\'{e}, 75005 Paris, France}

\date{\today}

\begin{abstract}
The recently introduced ``fuzzy sphere'' method has enabled accurate numerical regularizations of certain three-dimensional (3D) conformal field theories (CFTs). The regularization is provided by the non-commutative geometry of the lowest Landau level filled by electrons, such that the charge is trivially gapped due to the Pauli exclusion principle at filling factor $\nu=1$, while the electron spins encode the desired CFT. Successful applications of the fuzzy sphere to paradigmatic CFTs, such as the 3D Ising model, raise an important question: how finely tuned does the underlying electron system need to be? Here, we show that the 3D Ising CFT can also be realized at \emph{fractional} electron fillings. In such cases, the CFT spectrum is intertwined with the charge-neutral spectrum of the underlying fractional quantum Hall (FQH) state -- a feature that is trivially absent in the previously studied  $\nu=1$ case. Remarkably, we show that the mixing between the CFT spectrum and the FQH spectrum is strongly suppressed within the numerically-accessible system sizes. Moreover, we demonstrate that the CFT critical point is unaffected by the exchange statistics of the particles and by the nature of topological order in the charge sector. Our results set the stage for the fuzzy-sphere exploration of conformal critical points between topologically-ordered states. 
\end{abstract}

\maketitle

\section{Introduction}\label{sec:intro}
  
Understanding the universal properties of continuous phase transitions has been a long-standing area of focus~\cite{Sachdev11}. A powerful tool in this endeavor have been conformal field theories (CFTs)~\cite{DiFrancesco97} -- a class of interacting field theories with a rich symmetry structure that can emerge in statistical mechanics models tuned to a critical point~\cite{Polyakov70}. 
In two dimensions (2D), the symmetry is further enhanced, which leads to the exact solvability of such CFTs~\cite{Belavin84}, enabling tremendous progress in the analytical understanding of critical phenomena~\cite{Cardy1996}.

In contrast to this elegant framework, progress on CFTs in $d > 2$ dimensions has been hindered due to their different algebraic structure, requiring the use of sophisticated techniques such as conformal bootstrap~\cite{Poland19}. One promising direction for numerical studies of lattice models realizing CFTs is to use the so-called state-operator correspondence~\cite{Cardy1996}, which allows for the direct extraction of CFT data from a quantum Hamiltonian. This procedure, however, is exact only when the quantum theory is defined on $S^{d-1} \times \mathbb{R}$~\cite{Cardy84, Cardy85}, i.e., when the lattice model is embedded on a $(d{-}1)$-dimensional sphere $S^{d-1}$. For one-dimensional (1D) quantum systems, the ``sphere'' is simply a ring with periodic boundary conditions, which allows for treatment by standard numerical techniques~\cite{Henkel2013conformal,Lauchli13,Zou2018,Zou2020}. Unfortunately, for $d>2$, the $S^{d-1}$ manifold displays non-zero curvature and lattice models can no longer be seamlessly embedded into it.

\begin{figure}
    \centering
    \includegraphics[width=\linewidth]{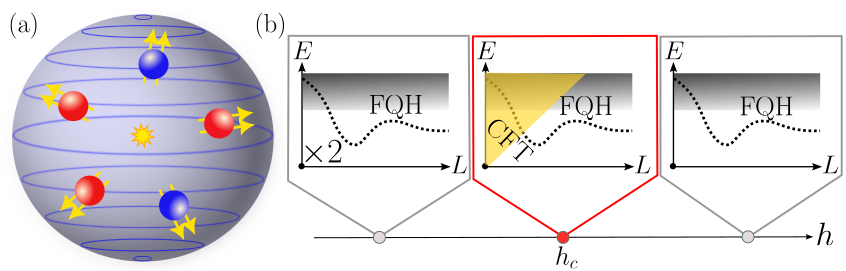}
    \caption{(a) Fuzzy sphere with a magnetic monopole at its center (star). In contrast to previous works that considered electrons as underlying degrees of freedom, we consider composite fermions, i.e., electrons dressed with two magnetic flux quanta (yellow arrows). The composite fermions carry an internal layer degree of freedom (depicted by blue and red). (b) A schematic of the phase diagram as a function of the tuning parameter $h$, with prototype spectra, energy $E$ vs. angular momentum $L$, in the gapped phases (gray boxes) and at the critical point (red box). In gapped phases, the low-lying spectrum consists of an FQH ground state and the gapped magnetoroton collective excitation (dashed line). In the ordered phase $h<h_c$, the ground state is two-fold degenerate, while it is unique in the disordered phase $h>h_c$, signalling an Ising-type transition. At the critical point, a gapless spectral branch described by CFT emerges (yellow triangle). }
    \label{fig:summary}
\end{figure}

In the special case $d=3$, this challenge has recently been circumvented in Ref.~\cite{Zhu23} by abandoning the lattice description and instead embedding the CFT into a continuum gas of electrons that fill the lowest Landau level (LLL) in a perpendicular magnetic field. The projection of the electrons' Hilbert space to the LLL results in non-commutativity of their coordinates, $[\hat X, \hat Y] = -i\ell_B^2$, where $\ell_B=\sqrt{\hbar/eB}$ is the magnetic length~\cite{Ezawa2008quantum}. 
This uncertainty relation smears the notion of a point and the electrons can be viewed as living on the surface of a ``fuzzy sphere''~\cite{Madore1992}, with the magnetic field generated by a Dirac monopole at its center~\cite{Haldane83}, see Fig.~\ref{fig:summary}(a). 

The numerics on the fuzzy sphere have successfully demonstrated conformal invariance and extracted CFT data in a number of Wilson-Fisher theories \cite{Zhu23, Han24, Han23, Hu23, Hofmann24, Hu24b, Zhou24c, Zhou24d, Dedushenko24, Cuomo24}. More recently, this scheme has been applied to deconfined quantum phase transitions in the $\mathrm{SO}(5)$ \cite{Zhou24a} and related $\mathrm{Sp}(N)$ theories \cite{Zhou24b}, showcasing its applicability to critical points both within the Landau-Ginzburg paradigm and beyond. In these applications, the electrons completely fill the LLL, forming an integer quantum Hall state.  However, a partially-filled LLL can host a multitude of exotic fractional quantum Hall (FQH) phases, such as Laughlin~\cite{Laughlin83}, composite fermion~\cite{Jain89} and Moore-Read~\cite{Moore91} states, in which electrons fractionalize into anyons~\cite{Arovas84}.   
Does the fuzzy sphere approach still work if the electrons form such complex many-body states? In other words, is the method applicable to the cases illustrated in Fig.~\ref{fig:summary}(a), where the underlying degrees of freedom are composite fermions, i.e., electrons dressed by an even number of magnetic flux quanta~\cite{Jain07}? 

In this paper, we argue that the fuzzy sphere regularization can indeed be applied to large classes of gapped FQH phases hosted in the charge sector. We illustrate this by realizing the 3D Ising CFT using several Abelian and non-Abelian FQH states. The key difference with previous work is that the low-energy spectrum of FQH states generically exhibits a gapped collective mode~\cite{Girvin85,Girvin86}, as observed in numerous experiments~\cite{Pinczuk93, Kang01, Kukushkin09}. In the long-wavelength limit, this mode is described by a variant of gravitational Chern-Simons theory whose formulation has attracted much effort over the past decade~\cite{Haldane11, Can14, Ferrari2014, Gromov14, Bradlyn15, You16, Gromov17, Nguyen18}. In the scenario explored here, the gapped FQH spectrum persists for all values of the tuning field $h$, see Fig.~\ref{fig:summary}(b), while precisely at criticality, the CFT branch  comes down and dominates the low-energy description. While we will show that there is only weak ``hybridization'' between the CFT and FQH spectrum in the examples considered below, the setup developed here may, in principle, allow to probe the intriguing possibility of strong coupling between such theories. 

The remainder of this paper is organized as follows. In \cref{sec:model,sec:ground_state} we introduce the model and investigate its ground state through exact diagonalization and mean-field theory. After presenting evidence for an Ising transition, we then assess the conformal symmetry at the critical point in \cref{sec:spectrum} through finite-size scaling and conformal perturbation. We observe good agreement between the spectrum of the model and the operator scaling dimensions of the 3D Ising CFT, even when such states are energetically near the charge-neutral excitations of the underlying FQH state. In \cref{sec:f-theorem} we demonstrate the decoupling between the charge and spin sectors of the model through the lens of F-theorem. We find that, once accounting for the interacting nature of the charge sector (by subtracting its topological entanglement entropy), the remaining entropy exhibits an almost identical behavior to that of the $\nu=1$ model. In the Appendices, we present additional data for the $\nu=1/3$ model and we demonstrate the broader applicability of our results to different filling factors and topological orders, with further details on the construction of minimal models.

\section{Model}\label{sec:model}

We employ the model of the fuzzy sphere similar to that of Ref.~\cite{Zhu23}. Consider $N$ particles on a sphere with radius $R$, with a magnetic monopole of strength $2Q$ placed at its center~\cite{Haldane83}. The particles carry an SU(2) internal degree of freedom which we can physically think of as a ``layer'' index, $\uparrow, \downarrow$. The monopole generates a radial magnetic field $\mathbf{B}=(2Q\phi_0/4\pi R^2)\hat{\mathbf{r}}$, where $\phi_0$ is the flux quantum. The magnetic field leads to the Landau level quantization for a particle on the surface of the sphere. Restricting to the Lowest Landau level (LLL) gives rise to an effective non-commutative geometry, where the notion of a point is not defined on length scales smaller than $\ell_B$. In the LLL, the single-particle states are given by monopole harmonics~\cite{Wu76,Wu77}. The monopole harmonics are approximately localized (to within $\sim\ell_B$) around the circles of latitude, see Fig.~\ref{fig:summary}(a), and there are $2Q+1$ such orbitals that are linearly independent and form the basis of  the LLL. The ratio of the number of particles and the degeneracy of the LLL allows us to define the filling factor $\nu$ in the thermodynamic limit, $\nu=\lim_{Q\to \infty}N/(2Q)$. 

Our model is described by the Hamiltonian
\begin{eqnarray}\label{eq:hamiltonian}
    H = H_\text{intra} + H_\text{inter} + H_t,    
\end{eqnarray}    
where the first two terms, corresponding to interactions between particles belonging to same (``intra'') or opposite layers (``inter''), produce an effective Ising coupling for the layer degrees of freedom, while the final term $H_t$ plays the role of a transverse field. Specifically, in the second-quantized form, the three terms are given by 
\begin{align}\label{eq:hamiltonian_terms}
    H_\text{intra} &= \sum_{a = \uparrow,\downarrow} \sum_{j_{1,2,3,4}=-Q}^Q V^\mathrm{intra}_{j_1j_2j_3j_4} (\mathbf{c}^\dag_{j_1} \sigma^a \mathbf{c}_{j_4}) (\mathbf{c}^\dag_{j_2} \sigma^a \mathbf{c}_{j_3}) \, , \\
    H_\text{inter} &= 2 \sum_{j_{1,2,3,4}=-Q}^QV^\mathrm{inter}_{j_1j_2j_3j_4} (\mathbf{c}^\dag_{j_1} \sigma^\uparrow \mathbf{c}_{j_4}) (\mathbf{c}^\dag_{j_2} \sigma^\downarrow \mathbf{c}_{j_3}) \, , \\
    H_t &= -h \sum_{j=-Q}^Q \mathbf{c}^\dag_j \sigma^x \mathbf{c}_j\, ,
\end{align}
where $\mathbf{c}_j = (c_{j\uparrow}, c_{j\downarrow})^\mathrm{T}$ is the bilayer annihilation operator for the $j$-th LLL orbital, and $\sigma^{\uparrow,\downarrow}$ are projectors onto the $\uparrow$ and $\downarrow$ components, respectively, and $h$ is the local magnetic field in the transverse $x$-direction which couples to the standard Pauli matrix $\sigma^x$. 

Due to the rotational invariance of the interactions, the Hamiltonian matrix elements $V_{j_1j_2j_3j_4}^\mathrm{intra}$, $V_{j_1j_2j_3j_4}^\mathrm{inter}$ are fully specified by a discrete set of numbers $\{ V_m\}$ called the Haldane pseudopotentials~\cite{Haldane83}, where integer $m$ represents the relative angular momentum of a pair of particles. For the illustrative example of $\nu=1/3$ state that we focus on throughout the main text, the pseudopotentials are chosen according to:
\begin{eqnarray}\label{eq:interaction}
\notag V^{\mathrm{intra}} &=& \{ V_0,V_1 \} = \{ 0,1 \}, \\
V^{\mathrm{inter}} &=& \{ V_{0},V_{1},V_{2},V_{3} \} = \{ 1,1,0.49,0.09 \}.
\end{eqnarray}
Note that the allowed values of pseudopotentials are constrained by the exchange statistics of the particles and their layer index. This is why, for example, we can set $V_0^\mathrm{intra}=0$ when we consider fermions~\cite{Prange87}. In Appendix~\ref{app:optimization} we provide further justification for the model in Eq.~\eqref{eq:interaction}, while the pseudopotentials for other filling factors are presented in Appendix~\ref{app:othernu}. 

In contrast to Ref.~\cite{Zhu23}, we will be interested in the cases $\nu < 1$ when the LLL is \emph{not} entirely filled by electrons. For the Laughlin $\nu=1/3$ state~\cite{Laughlin83} considered below, we set the monopole strength according to $2Q = 3N-3$, with the constant offset representing the Wen-Zee shift \cite{WenZee}. Another difference with respect to Ref.~\cite{Zhu23} is that we require the presence of both intra- and inter-layer interactions in our Hamiltonian. Furthermore, as we have moved away from integer filling, the microscopic particle-hole symmetry is no longer present in our case. Instead, the principal microscopic symmetry that we will exploit is the Ising $\mathbb{Z}_2$ symmetry, which relates the two layers by $\mathbf{c}_j \to \sigma^x \mathbf{c}_j$. Finally, unlike Ref.~\cite{Zhu23} which only considered particles with fermionic statistics, in Appendix~\ref{app:othernu} we will show that our results also hold for bosonic particles with an appropriate interaction. 
 
\section{Ground state}\label{sec:ground_state}

In the $\nu=1$ case, when the spin is fully polarized by the transverse magnetic field, the ground state is necessarily gapped due to the Pauli exclusion principle: it is not possible to add another electron (with the same spin) to the already filled LLL. For fractional $\nu$, the behavior of the charge sector is far more complicated, even if the spin is fully polarized: depending on the filling and the interaction potential, a large variety of phases can be realized, including different gapped and compressible FQH states, such as the composite fermion Fermi liquid and symmetry-broken phases~\cite{JainHalperin2020}.  

In this section, using a combination of exact diagonalization (ED) and mean-field (MF) theory, we argue that the ground state of the model in Eq.~\eqref{eq:interaction} at filling factor $\nu=1/3$ and for all values of the transverse field $h$ is described by the Laughlin state~\cite{Laughlin83}:
\begin{eqnarray}\label{eq:laughlin}
    \psi_{\nu=1/q}(\{u_i,v_i\}) = \prod_{i<j}^N (u_i v_j-u_j v_i)^q,
\end{eqnarray}
which has been written in terms of standard spinor coordinates $u_j = \cos(\theta_j/2)\exp(i\phi_j/2)$, $v_j = \sin(\theta_j/2)\exp(-i\phi_j/2)$ on the fuzzy sphere~\cite{Haldane83}. The wave function in Eq.~\eqref{eq:laughlin} describes only the orbital state of the electrons, i.e., the full wave function is given by a tensor product $|\Psi\rangle = |\psi_{\nu=1/q}\rangle \otimes |\chi\rangle$, where $|\chi\rangle$ describes the spin component of the wave function. The latter depends on the value of the transverse field $h$, as we discuss below. Note that, if $|\chi\rangle$ does not have maximal spin polarization, the full wave function $|\Psi\rangle$ needs to be explicitly (anti)symmetrized to make it consistent with the exchange statistics of the particles. 

\begin{figure}[t]
    \centering
    \includegraphics[width=0.48\textwidth]{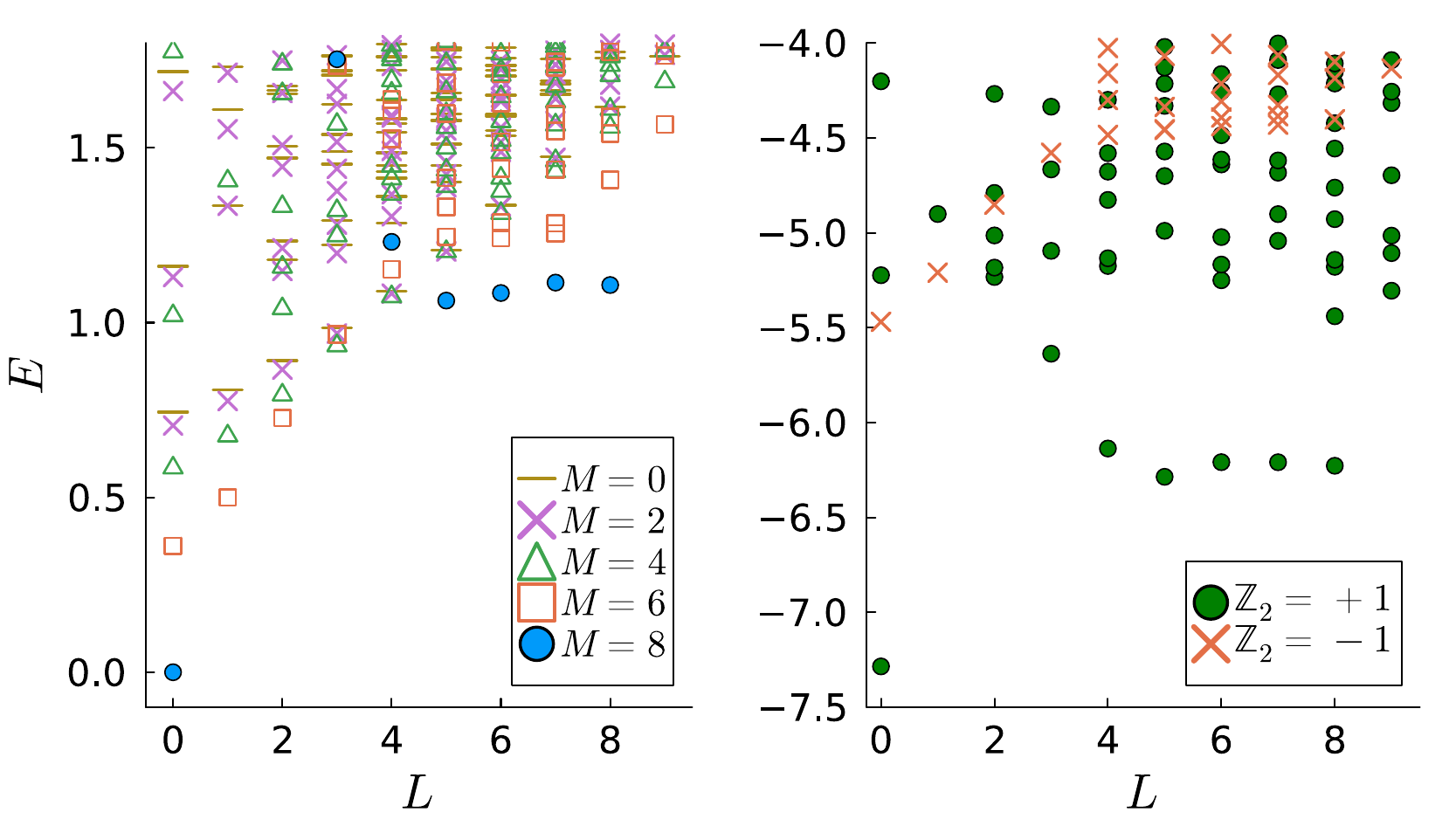}
    \caption{Energy spectrum of the bilayer model in Eq.~\eqref{eq:interaction} plotted as a function of angular momentum $L$. All data is for $N=8$ particles at filling $\nu=1/3$, obtained by ED. Left panel: Ferromagnetic phase at $h=0$, where the magnetization $M$, Eq.~\eqref{eq:magnetization}, is a good quantum number that has been resolved. The spectrum is invariant under $M \to -M$, and only $M \geq 0$ is shown.  Thus, there are two degenerate ground states with $E=0$ (one at $M = 8$ and one at $M=-8$), which are the \emph{exact} Laughlin states. The corresponding magnetoroton branch in the fully layer-polarized part of the spectrum can be observed, although it is partially masked by other low-lying excitations corresponding to spin waves (e.g., in sector $M=N-2$). Right panel: Paramagnetic phase at $h=1$. There is now a unique Laughlin ground state, with the magnetoroton branch present in the even-parity sector.}
    \label{fig:ferromagnet+paramagnet}
\end{figure}

\subsection{Exact diagonalization}\label{sec:groundstateED}

For our interaction in Eq.~(\ref{eq:interaction}), the existence of the Laughlin state in the $h\gg 1$ limit is ensured by the dominance of $V_1$ pseudopotential. Namely, the large $x$-field polarizes the system and the resulting single-layer system is described by an effective ``symmetrized'' interaction, $(V^\mathrm{intra}+V^\mathrm{inter})/2$~\cite{Papic10}. Consequently, in that limit, the even pseudopotentials $V_0$, $V_2$ in Eq.~(\ref{eq:interaction}) drop out, while the $V_1$ pseudopotential -- the parent Hamiltonian of the $\nu = 1/3$ Laughlin state~\cite{Haldane83} -- remains as the leading contribution to the interaction. This analysis furthermore predicts that the entire low-energy spectrum at $h\gg 1$ should correspond to that of the Laughlin state, with the characteristic gapped collective mode known as the magnetoroton or the Girvin, MacDonald and Platzman mode~\cite{Girvin85,Girvin86}. This is confirmed in \cref{fig:ferromagnet+paramagnet}.

\begin{figure}[t]
    \centering
    \includegraphics[width=0.48\textwidth]{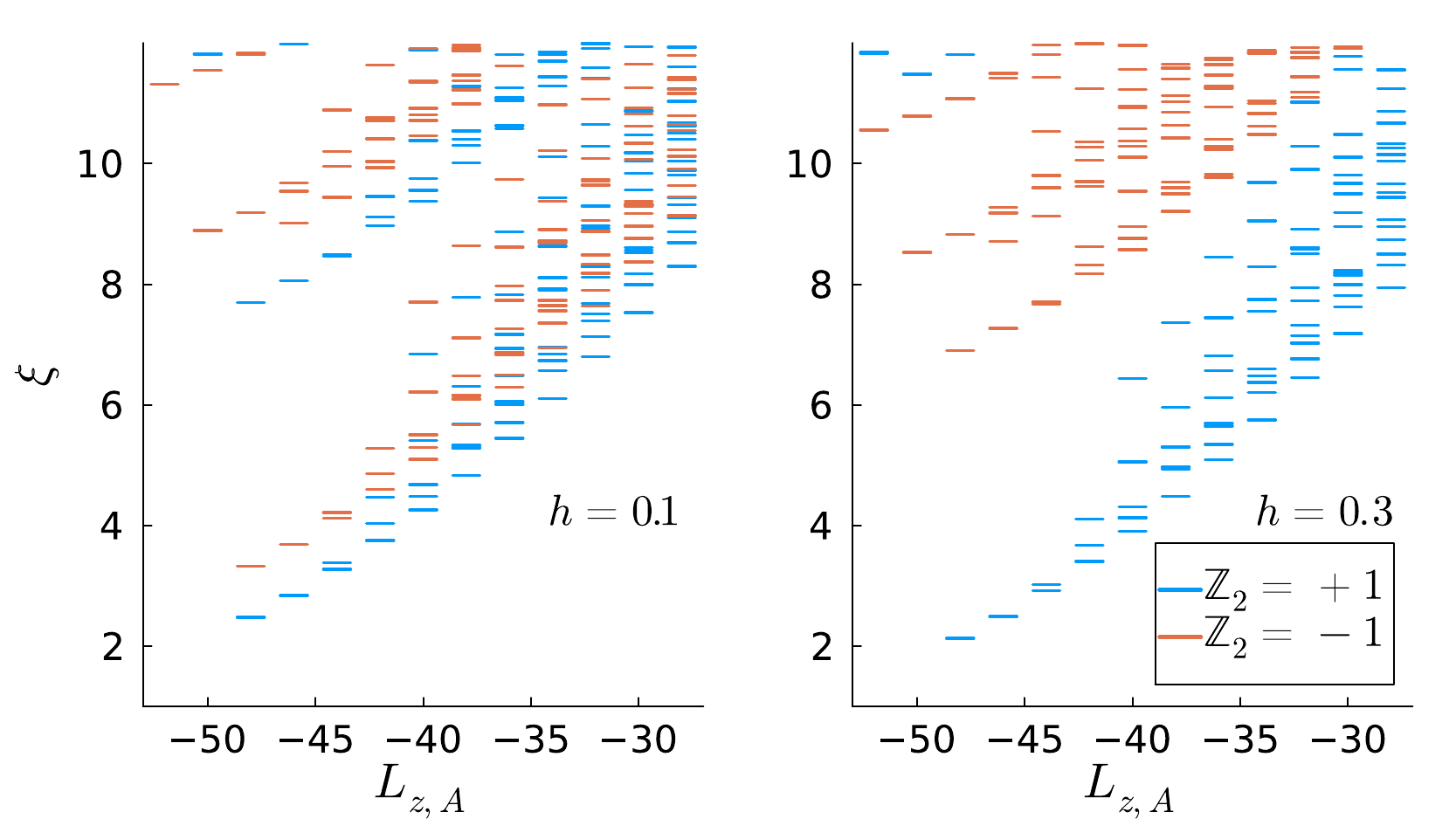}
    \caption{Real-space entanglement spectrum of the ground state of interaction~\eqref{eq:interaction} in the ferromagnetic ($h=0.1$) and paramagnetic ($h=0.3$) regimes. Data is for $N=8$ particles, in the $N_A = 4$ sector. The state counting in both $\mathbb{Z}_2$ sectors is $1,1,2,3,5,\dots$, following that of a chiral boson edge mode, while the entanglement gap between the two sectors increases with the transverse field.}
    \label{fig:RSES}
\end{figure}

To obtain an Ising-type transition, we need two degenerate ferromagnetic ground states in the limit of no transverse field ($h=0$). Furthermore, those ground states should still belong to the same Laughlin phase or else we may have another phase transition in the charge sector, which would severely complicate the analysis. In \cref{fig:ferromagnet+paramagnet} we confirm that, at $h=0$, the ground state is fully spin-polarized by evaluating the expectation value of magnetization (layer polarization),
\begin{equation}\label{eq:magnetization}
    M = \sum_{j=-Q}^Q \mathbf{c}^\dagger_j \sigma^z \mathbf{c}_j \, ,
\end{equation}
which is a good quantum number in the absence of $h$. As expected, the ground state is two-fold degenerate, $M = \pm N$. For general values of $h$, the ground state is identified with the Laughlin state by examining the counting of the entanglement spectrum, which matches the counting of edge modes of a chiral boson~\cite{Li08} -- see Fig.~\ref{fig:RSES}.

\begin{figure}[tb]
    \centering
    \includegraphics[width=0.48\textwidth]{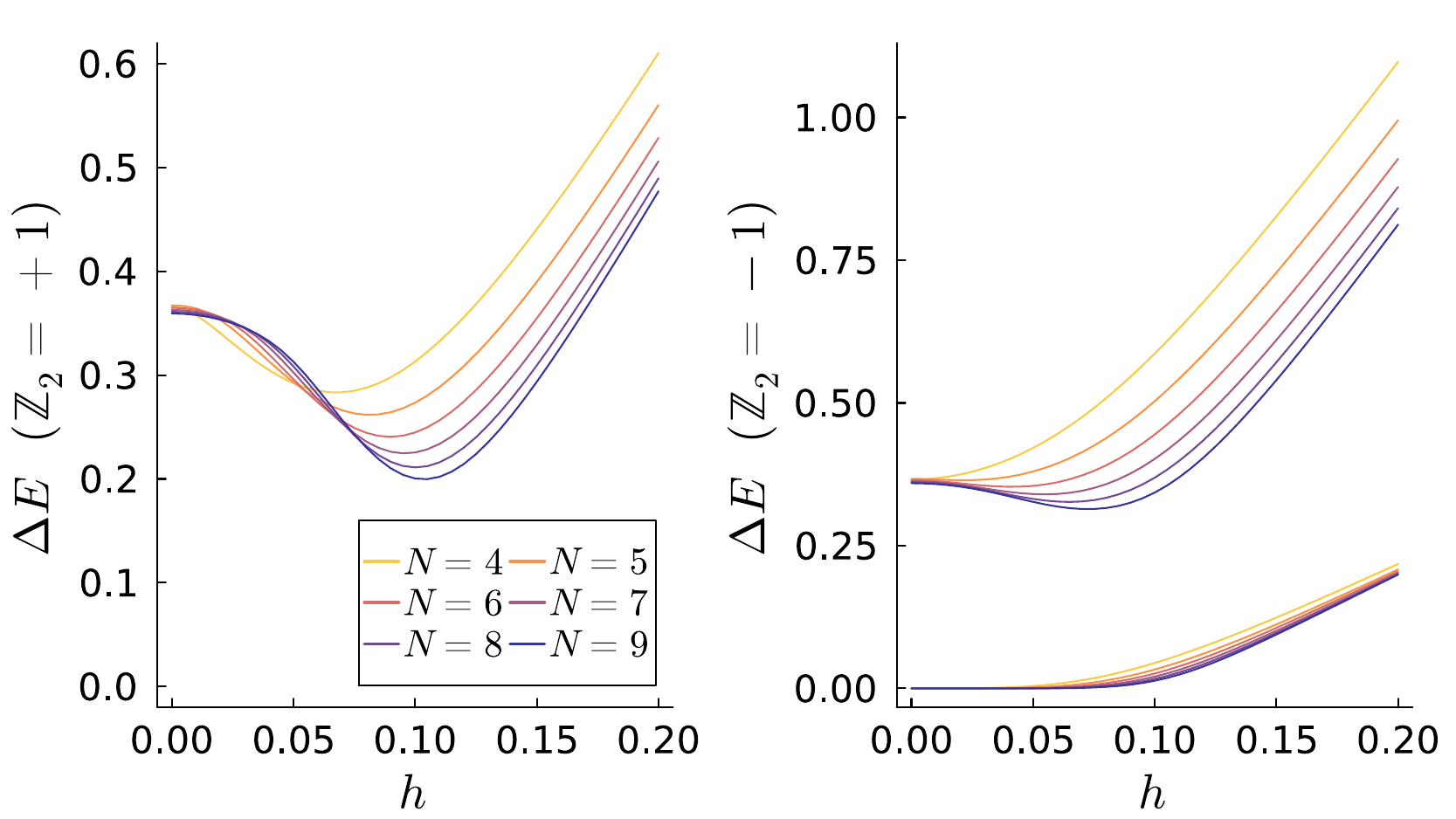}
    \caption{Gaps to the first excited states above the ground state in the $\mathbb{Z}_2$-even (left panel) and $\mathbb{Z}_2$-odd (right panel) sectors. The even sector is gapped on either side of the transition. In the odd sector, we can observe the ground state degeneracy being lifted across the transition, while the next excited state also has a finite gap in the ferromagnetic regime. The decrease (and subsequent closing in the thermodynamic limit) of the shown gaps is consistent with a continuous phase transition around $h \sim 0.1$.}
    \label{fig:gaps}
\end{figure}

The system in both $h=0$ and large-$h$ limits exhibits a finite energy gap that quickly converges with respect to the system size, as demonstrated in \cref{fig:gaps}. By contrast, around $h\sim 0.1$, the gap in both $\mathbb{Z}_2$ symmetry sectors is seen to be closing with system size, indicative of a continuous phase transition -- see Appendix~\ref{app:gaps} for further details. This critical point will be analyzed more carefully in~\cref{sec:spectrum} below.

\subsection{Mean-field approximation}\label{sec:mean-field}

To understand the nature of the ground state at intermediate values of the field $h$, we apply a MF treatment {\it in the spin sector}. We approximate the ground state of the bilayer model with a Laughlin state (or a state close by), rotated by an optimal angle $\theta_\text{opt}$. Numerically, such a MF state can be constructed as follows~\cite{Thiebaut2014}. We project the fermionic operators onto any chosen direction $\theta$:
\begin{equation}\label{eq:projection1}
    \mathbf{c}_j = (c_{j\uparrow}, c_{j\downarrow})^T \to \tilde{\mathbf{c}}_j = (\cos \frac{\theta}{2} \, c_{j\uparrow} + \sin \frac{\theta}{2} \, c_{j\downarrow}, 0)^T . 
\end{equation}
Bilayer density terms, defined by an arbitrary spin matrix $\sigma^a$ acting in the spin space, take the form
\begin{equation}\label{eq:projection2}
    \mathbf{c}^\dag_{j_1} \sigma^a \mathbf{c}_{j_2} \to \tilde{\mathbf{c}}^\dag_{j_1} \left( \mathcal{R}_{\theta/2}  \, \sigma^a \, \mathcal{R}^T_{\theta/2}  \right)  \tilde{\mathbf{c}}_{j_2} ,
\end{equation}
where $\mathcal{R}_{\theta/2} $ is the rotation matrix. This similarly applies to the density-density interactions in \cref{eq:hamiltonian_terms}, from which we can extract the effective pseudopotentials:
\begin{equation}\label{eq:mfinteraction}
    \tilde{V}_m = \left(\cos^4 \frac{\theta}{2} + \sin^4 \frac{\theta}{2}\right) V_m^\text{intra} + 2\cos^2 \frac{\theta}{2} \sin^2 \frac{\theta}{2} \, V_m^\text{inter} \, .
\end{equation}
Upon diagonalizing this single-layer interaction, we obtain the MF state $|\psi_\text{MF}(\theta)\rangle$. It is important to note that while this treatment discards correlations in the spin sector, it preserves the interacting nature of the charge sector. The effective Hamiltonian also preserves the $\mathbb{Z}_2$ Ising symmetry, as it is invariant under $\theta \to \pi - \theta$.

\begin{figure}[tb]
    \centering
    \includegraphics[width=0.48\textwidth]{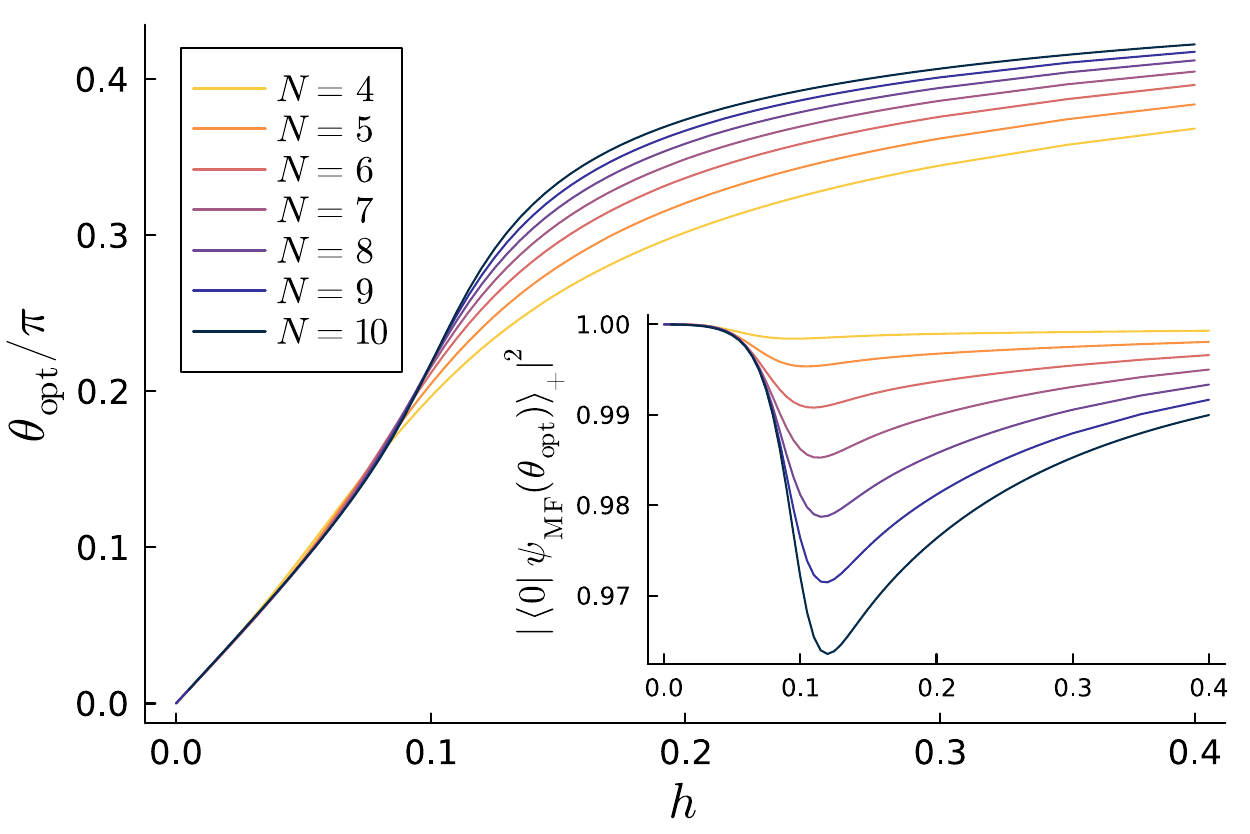}
    \caption{Mean-field approximation for the bilayer Laughlin ground state. The polarization $\theta_\text{opt}$ is found by maximizing the overlap of the MF state, defined via Eq.~\eqref{eq:mfinteraction}, with the bilayer ground state using gradient descent (inset). In the ferromagnetic limit, the ground state is fully layer-polarized, while in the paramagnetic regime it approaches full transverse polarization. As expected, the MF approach worsens with increased system size at the critical point. }
    \label{fig:mean-field}
\end{figure}

The optimal angle $\theta_\text{opt}$ is chosen such that the bilayer state $|\psi_\text{MF}(\theta_\text{opt})\rangle_{+} = \left( |\psi_\text{MF}(\theta_\text{opt})\rangle + |\psi_\text{MF}(\pi-\theta_\text{opt})\rangle \right)/\sqrt{2}$ maximizes overlap with the ground state of the bilayer system; note that this approach can also be used to automatically construct the lowest state in the $\mathbb{Z}_2$-odd sector, $|\psi_\text{MF}(\theta_\text{opt})\rangle_{-} = \left( |\psi_\text{MF}(\theta_\text{opt})\rangle - |\psi_\text{MF}(\pi-\theta_\text{opt})\rangle \right)/\sqrt{2}$. This approximation has a twofold benefit. First, the evolution of the effective polarization offers a simple physical intuition for the phase transition, as shown in \cref{fig:mean-field}. The ground state starts out as effectively layer-polarized, and gradually tilts towards $\theta = \pi/2$ in the paramagnetic phase. However, the overlap shows a characteristic dip in the vicinity of the critical point, where the spin correlations become significant. In addition, the MF picture can improve our understanding of the underlying field theory, and will be applied to the study of state-operator correspondence and entanglement entropy, in \cref{sec:spectrum,sec:f-theorem} respectively.

In a similar manner, we can construct MF approximations to the magnetoroton states by embedding  single-layer states into the bilayer system. In the long-wavelength limit, the states are well described by the single-mode approximation (SMA)~\cite{Girvin86}: 
   \begin{align}\label{eq:sma}
   |\phi_L^\mathrm{SMA} \rangle = &\frac{1}{\sqrt{2}} \big( \tilde{\rho}_{L,0}(\theta_\text{opt}) |\psi_\text{MF} (\theta_\text{opt})\rangle \nonumber \\ &\pm \tilde{\rho}_{L,0}(\pi -\theta_\text{opt}) |\psi_\text{MF} (\pi - \theta_\text{opt})\rangle \big) \, ,
\end{align}
where $\tilde{\rho}_{L,0} $ is the $(L,0)$ component of the Fourier transform of the single-layer density operator $\tilde{\rho}$. This construction will be useful for identifying energy levels that form part of the non-CFT magnetoroton branch in the following section. Note that the different sign will allows us to construct magnetoroton modes in both parity sectors in the following section.

\section{Ising conformal critical point}\label{sec:spectrum}

Thus far, we have established that the ground state changes from a $\mathbb{Z}_2$-ferromagnet to a paramagnet as the transverse field is increased, hence we expect a continuous Ising-type transition along the way. In this section we demonstrate that the critical point falls in the universality class of the 3D Ising CFT, and one can use radial quantization on the fuzzy sphere to make predictions about the spectrum and critical exponents at the phase transition.

\begin{figure}[tb]
    \centering
    \includegraphics[width=0.48\textwidth]{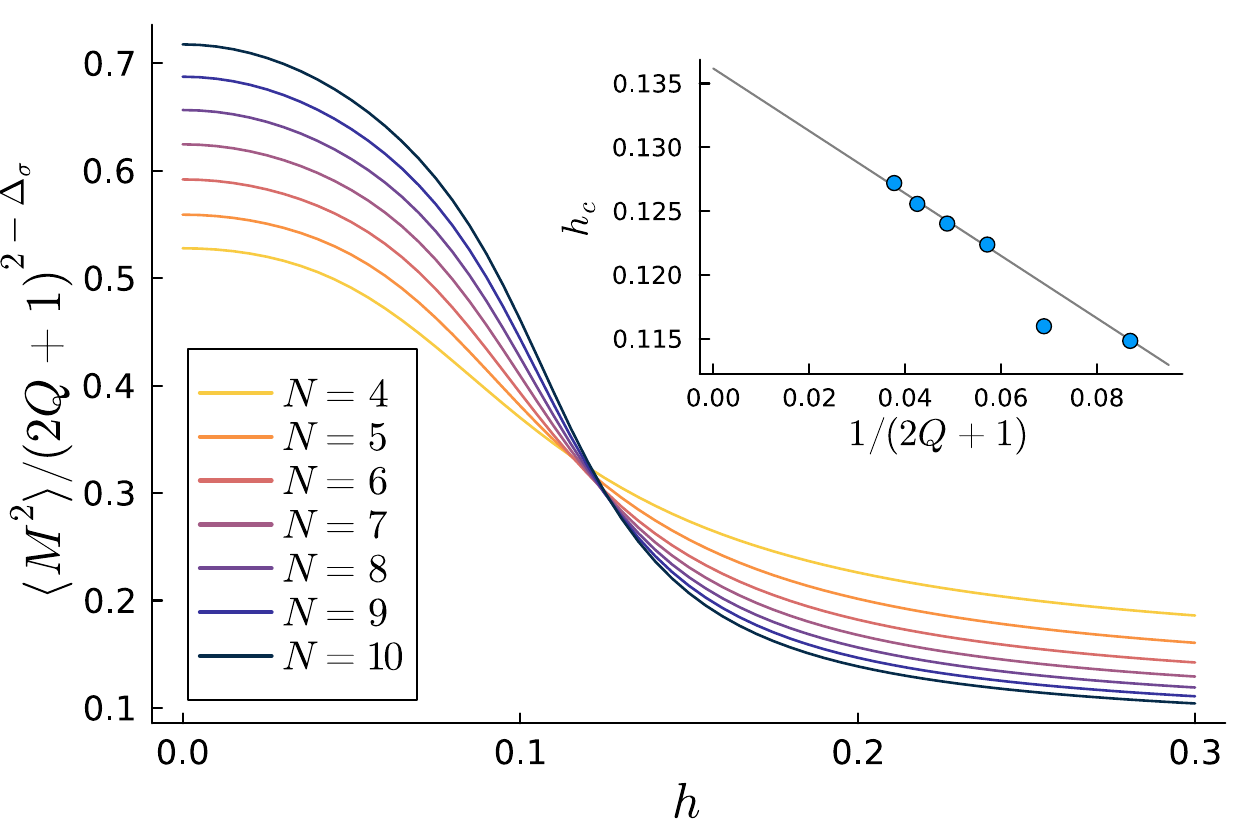}
    \caption{Finite size scaling of the order parameter in the ground state of the model in Eq.~\eqref{eq:interaction}. Assuming the scaling dimension $\Delta \approx 0.5181489$~\cite{Simmons-Duffin2017}, the rescaled data for system sizes $N>5$ show an approximate collapse. By extrapolating the crossings of the finite-size pairs ($N$, $N+1$), as shown in the inset, we extract the critical field in the thermodynamic limit at $h_c \approx 0.135$.}
    \label{fig:order_parameter}
\end{figure}

\subsection{Finite size scaling and state-operator correspondence}\label{sec:scaling}

Before attempting to match CFT operators with energy levels of finite systems, we first need to locate the critical point as accurately as possible. We do this with the help of magnetization in \cref{eq:magnetization} as the order parameter.
The critical point is identified as the crossing in the quantity $\langle M^2 \rangle / (2Q+1)^{2-\Delta_\sigma}$, where we use $\sqrt{2Q+1}$ as the length scale of the model (i.e., the radius of the fuzzy sphere), and $\Delta_\sigma = 0.5181489$ is the scaling dimension of the $\sigma$ primary operator~\cite{Simmons-Duffin2017}. The scaling of the order parameter is shown in \cref{fig:order_parameter}; by extrapolating the finite-size crossings, we identify the critical point at $h_c \approx0.135$. Note that the fractional filling considerably increases the dimension of the Hilbert space, limiting the exact diagonalization to $N \lesssim 10$ particles; the largest Hilbert space used has dimension $149674426$.
 
State-operator correspondence \cite{Cardy84,Cardy85} can be achieved by foliating $\mathbb{R}^3$ into $S^2 \times \mathbb{R}$, and defining the Hilbert space -- the physical space of the electrons -- on each leaf of foliation $S^2$. Different leaves are connected by the time evolution operator $U = \exp (-D \tau)$, hence the Hamiltonian is equal to the dilation generator $D$. Since the dilation and the Lorentz spin generators commute, we can classify the states of the spectrum according to their scaling dimension $\Delta$ and spin $L$, in one-to-one correspondence with the operators of the CFT.

The spectrum at the critical point consists of conformal towers, emerging from the primary operators of the theory. By definition, these operators are annihilated by the generator of special conformal transformations, which has been explicitly verified for the Ising point on the fuzzy sphere at $\nu=1$~\cite{Fardelli24,Fan24}. For any such primary $\mathcal{O}$ of dimension $\Delta$ and spin $L$, we can construct an infinite number of descendants of the form $\partial_{\nu_1} \dots \partial_{\nu_j} \partial_{\mu_1} \dots \partial_{\mu_i} \square^n \mathcal{O}_{\mu_1 \mu_2 \dots \mu_L}$, of dimension $\Delta + 2n + i + j$ and spin $L - i + j$, where $i \leq L$ and $\square$ denotes the Laplace operator.

\begin{figure}[tb]
    \centering \includegraphics[width=0.48\textwidth]{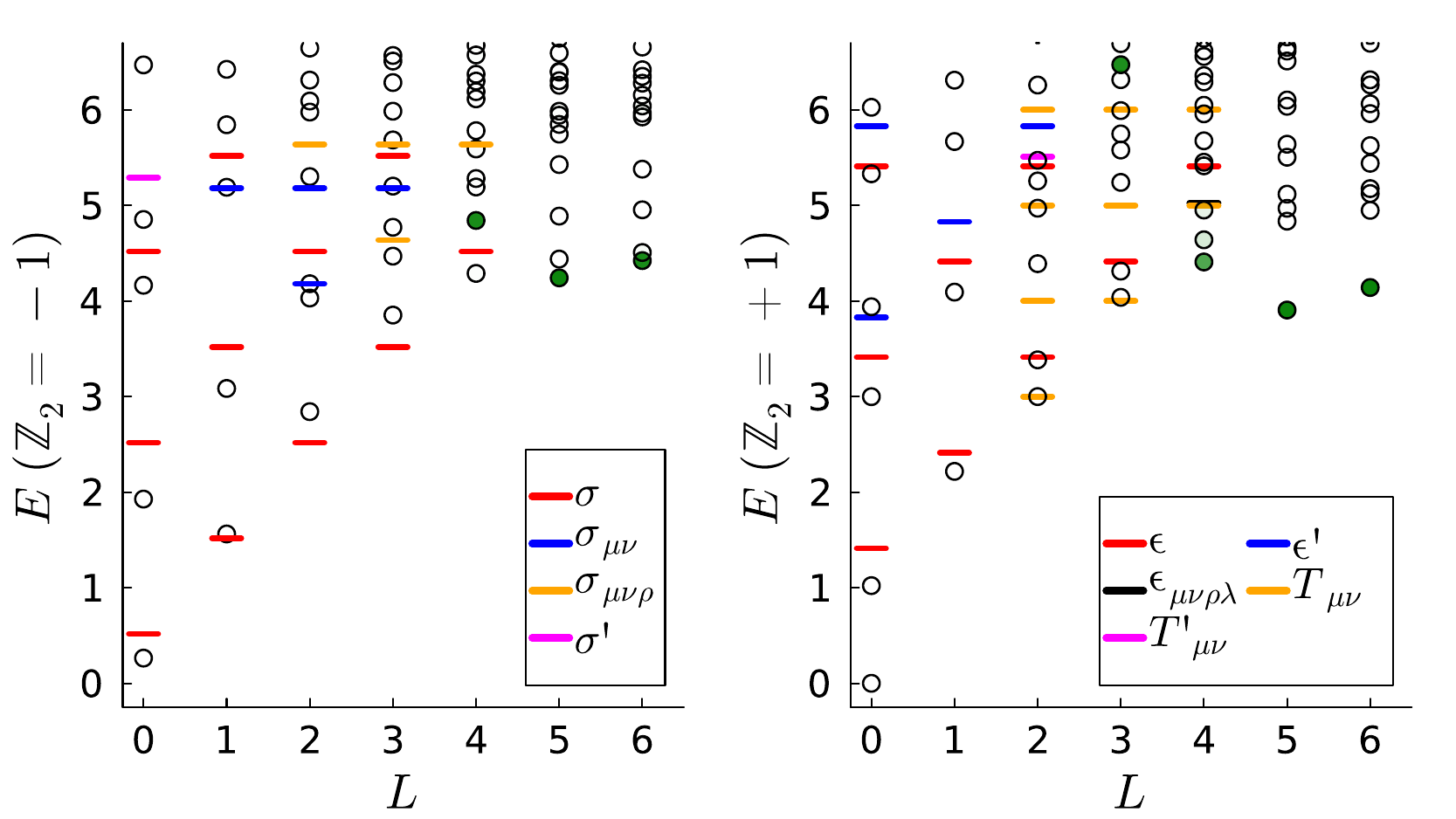}
    \caption{Energy spectrum of the model in Eq.~\eqref{eq:interaction} with $N=8$ at the critical value $h_c = 0.135$, estimated in Fig.~\ref{fig:order_parameter}. Empty circles are ED data and have been rescaled such that the energy-momentum  tensor (i.e., the lowest state with an even parity and Lorentz spin $L=2$) is at $E_T = 3$~\cite{Zhu23}. The lines are scaling dimensions (obtained by conformal bootstrap) corresponding to towers of CFT operators in the legend. Filled green circles mark the magnetoroton states, with their color intensity determined by the overlap with the corresponding MF state. In the $L=3$ sector, the MF magnetoroton state (shown in the $\mathbb{Z}_2$-even sector) sits inside the continuum but is still sharply peaked over a single eigenstate, with overlap 0.89; for the $L=4$ sector, the even-parity sector splits over the three lowest eigenstates with overlaps 0.69, 0.16, 0.09, while in the odd-parity sector it has a single peak of 0.90. In higher momentum sectors, e.g., $L=5,6$, the magnetoroton is a sharp excitation (overlap $> 0.90$ with a single eigenstate), while for $L=2$ it resides deeply in the spectral continuum beyond the scale of this figure.}
    \label{fig:conformal_spectrum_hc_inf}
\end{figure}

In \cref{fig:conformal_spectrum_hc_inf}, we show the spectrum at the identified critical point $h_c = 0.135$ for $N=8$ particles. After accounting for the speed of light, the broad features of the spectrum resemble the expected CFT spectrum, albeit there remain visible quantitative differences between the two. One could envision two possible explanations for this discrepancy: either the CFT spectrum is perturbed due to the small finite size of the system and less-than-optimal interaction parameters, or there is a mixing with the charge-neutral FQH excitations that renormalizes the energies. The FQH magnetoroton can be seen at large momenta in both parity sectors of \cref{fig:conformal_spectrum_hc_inf}. 
To identify the spinless excitations that affect our finite-size CFT spectrum, we have applied the MF treatment of \cref{sec:mean-field}, by checking overlaps with the magnetoroton states of the effective single-layer problem (which, at small $L$, are identified using the SMA). At sufficiently large $L\geq 5$, the magnetoroton states in \cref{fig:conformal_spectrum_hc_inf} can be unambiguously identified as having large overlap (exceeding 0.90) with a single eigenstate. However, at smaller momenta where the magnetoroton enters the continuum of the spectrum, e.g., $L=3,4$, we see that the MF approximation can have overlap over multiple eigenstates. 

In the next subsection, we will show that the mixing with FQH spectrum does not have a major influence on the CFT spectrum by deforming the model in accordance with conformal perturbation theory. In the thermodynamic limit, the lack of mixing may be anticipated due to the fact that the magnetoroton branch remains gapped throughout the transition, hence it will be pushed upwards in energy as the system size is increased -- this can be seen clearly in \cref{app:gaps}, and is due to the normalization imposed on the spectrum by fixing the energy-momentum tensor to $E_T=3$, while in absolute units $E_T$ decreases as $\sim 1/R$. However, the very weak mixing between the CFT and FQH spectra in small finite systems, where their energies are of the same order, is unexpected.

\subsection{Conformal perturbation}

The microscopic Hamiltonian $H$ near the critical point can be interpreted as the ``pure'' CFT Hamiltonian which is perturbed by integrals of the CFT operators:
\begin{equation}\label{eq:conformal_perturbation}
    H = \frac{1}{\alpha} \left( H_\text{CFT} + \sum_\mathcal{V} g_\mathcal{V} \int \mathrm{d}^2\Omega  \, \mathcal{V}(\Omega) \right) \, , 
\end{equation}
where $\mathcal{V}$ are $\mathbb{Z}_2$-even primaries, and $g_\mathcal{V}$ are their respective couplings. 
This approach, known as conformal perturbation theory~\cite{ZamolodchikovCtheorem}, was recently applied to the 3D Ising CFT realized by a small system of spins arranged on an icosahedron~\cite{Lao23}.
 We shall consider only the relevant operator $\epsilon$, and the first irrelevant operator $\epsilon'$, as they are enough to capture most of finite-size effects. We will study their effect on the energy levels $\sigma,\epsilon,\partial\sigma, \partial\epsilon$.

A given energy level of the microscopic model will be proportional to the scaling dimension of the corresponding operator $\Delta_\mathcal{O}$, corrected by an energy $\delta E^\mathcal{V}_\mathcal{O}$ that depends on both the perturbation and the state. In other words,
\begin{equation}
    E_\mathcal{O} = \frac{1}{\alpha} \left( \Delta_\mathcal{O} + \delta E_\mathcal{O}^{(\epsilon)} + \delta E_\mathcal{O}^{(\epsilon')}\right) \, ,
\end{equation}
where $1/\alpha$ is the speed of light of the model, and $\delta E^\mathcal{V}_\mathcal{O} = g_\mathcal{V} f_{\mathcal{O}\mathcal{V}\mathcal{O}}$ (for primaries, with slightly different form for descendants \cite{Lao23}). We assume that the OPE coefficients $f_{\mathcal{O}\epsilon\mathcal{O}}$ and $f_{\mathcal{O}\epsilon'\mathcal{O}}$ are known (e.g., using the values obtained by conformal bootstrap in \cite{Simmons-Duffin2017}), and we minimize 
\begin{equation}
    \delta = \sum_{\mathcal{O} \in \{\sigma, \partial\sigma, \epsilon, \partial\epsilon \}} \left(  \alpha E_\mathcal{O} - \Delta_\mathcal{O} - \delta E_\mathcal{O}^{(\epsilon)} - \delta E_\mathcal{O}^{(\epsilon')} \right)^{2}
\end{equation}
over the chosen operators to find the couplings $g_{\epsilon},g_{\epsilon'}$.

\begin{figure}[tb]
    \centering
    \includegraphics[width=0.48\textwidth]{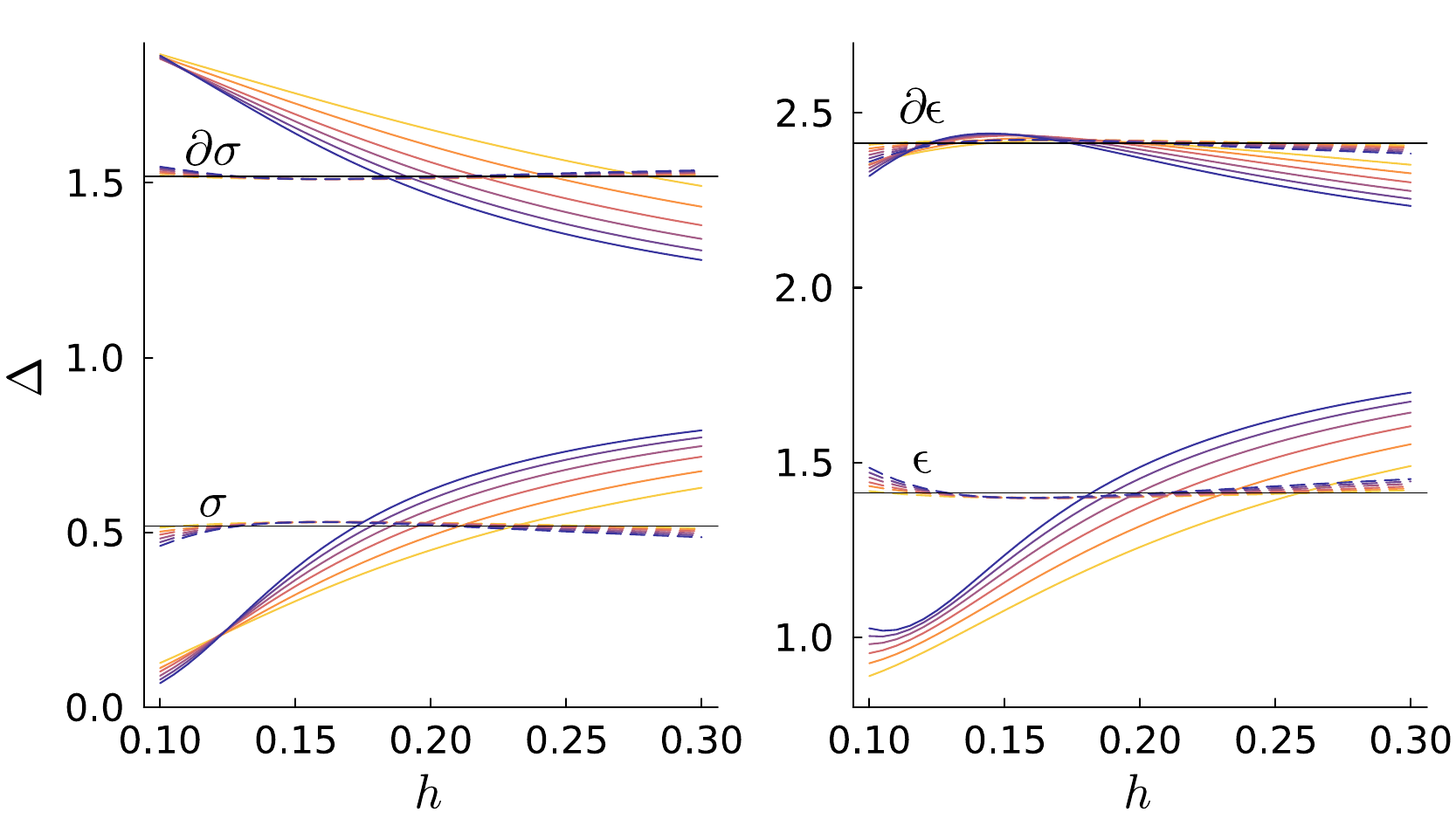}
    \caption{The evolution of energy levels $\sigma, \epsilon, \partial \sigma, \partial \epsilon$ across the transition for $N=4-9$ particles (colors represent different system sizes, with darker colors corresponding to larger $N$). Solid lines are the raw values (after a single-parameter fit using the speed of light $1/\alpha$), dashed lines are corrected values using the $\epsilon$ and $\epsilon'$ perturbations, and horizontal lines are predictions based on conformal bootstrap. }
    \label{fig:conformal_perturbation}
\end{figure}

Figure~\ref{fig:conformal_perturbation} shows the uncorrected energies, $\alpha E_{\mathcal{O}}$, alongside the corrected ones, $\big(\alpha E_{\mathcal{O}} - \delta E_{\mathcal{O}}^{(\epsilon)} - \delta E_{\mathcal{O}}^{(\epsilon')}\big)$, across  the transition. The application of the conformal perturbation visibly improves the matching between the ED data and the theoretical values. 

\begin{figure}[tb]
    \centering
    \includegraphics[width=0.48\textwidth]{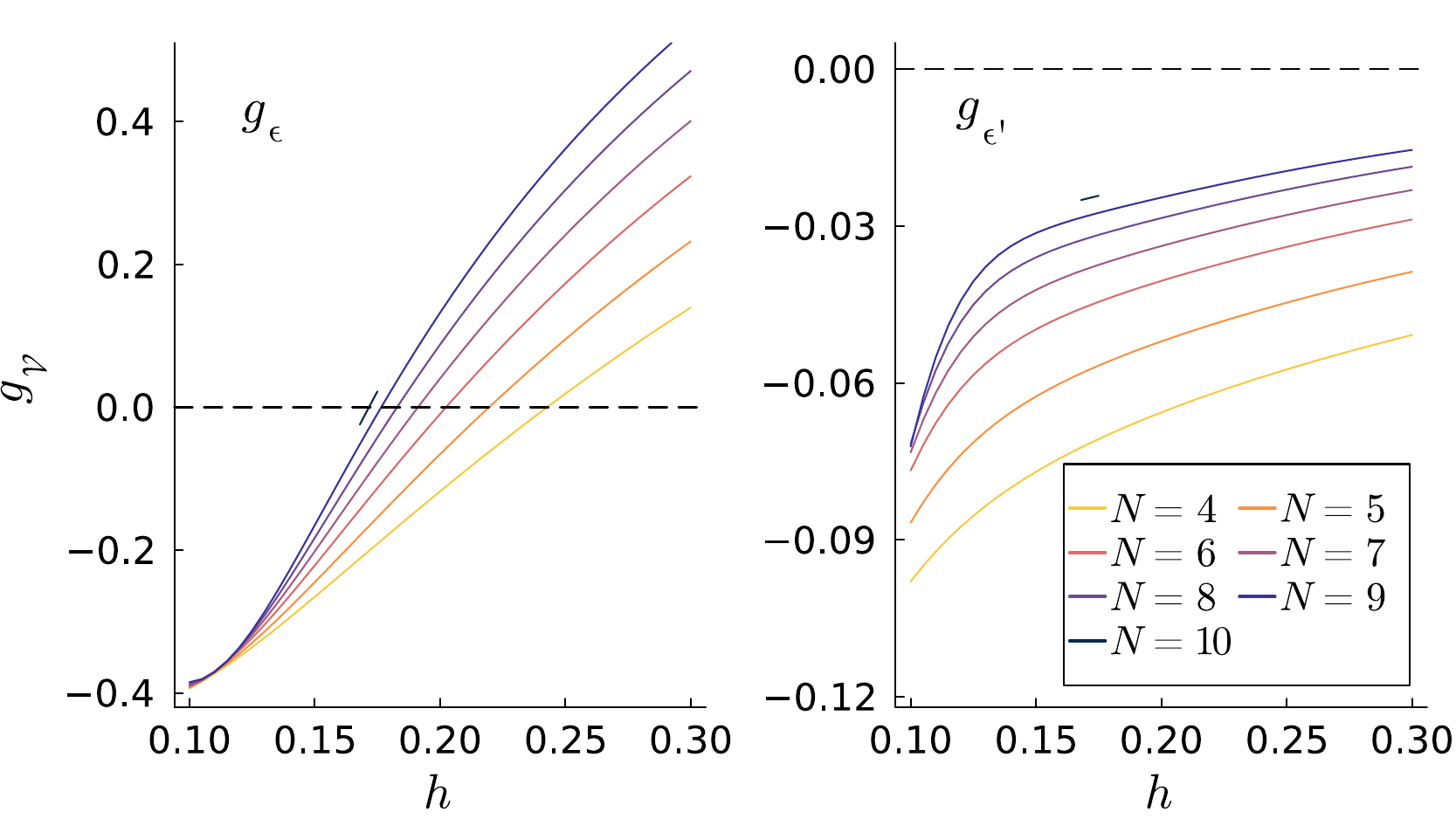}
    \caption{Coupling of the $\epsilon$ and $\epsilon'$ perturbations as a function of system size and transverse field. For the relevant $\epsilon$, the coupling needs to be zero at the phase transition, allowing us to identify the intercepts with the x-axis as the finite-size critical transverse fields $h_c(Q)$. For the largest system size $N=10$, we have only collected data in the vicinity of the crossing.
    A linear extrapolation in $1/(2Q+1)$ gives a critical point of $h_{c} \approx 0.13$ -- close to the value predicted by the order parameter. In contrast, the $\epsilon'$ coupling shows behavior indicative of an irrelevant operator, as its magnitude scales inversely with the system size.}
    \label{fig:coupling_epsilon}
\end{figure}

Figure~\ref{fig:coupling_epsilon} shows the coupling of the $\epsilon$ and $\epsilon^\prime$ perturbations. The latter is an irrelevant operator and, indeed, we observe that the coupling associated with it monotonically decreases with system size. On the other hand, $\epsilon$ is a relevant operator, hence the transition can be located as the point $h_c$ where its coupling goes to zero, $g_\epsilon(h_c(Q)) = 0$. Our model exhibits a strong dependence of the critical field on the particle number; extrapolating the value of $h_c(Q)$, we obtain the critical value of the field in the thermodynamic limit  $h_c \approx 0.13$ [linear extrapolation in $1/(2Q+1)$]. This is close to the previously determined crossing in the order parameter, serving as an important consistency check for our estimate of the transition. However, it is worth emphasizing that being able to determine the finite-size critical fields $h_c(Q)$ is, in fact, one of the strengths of the fuzzy sphere regularization. This allows us to extract CFT data from system sizes as small as $N=4$ electrons. We shall use the knowledge of $h_c(Q)$ in the next section to further demonstrate agreement between the entanglement properties of integer and fractionally filled models. 

\begin{figure}[tb]
    \centering 
    \includegraphics[width=0.5\textwidth]{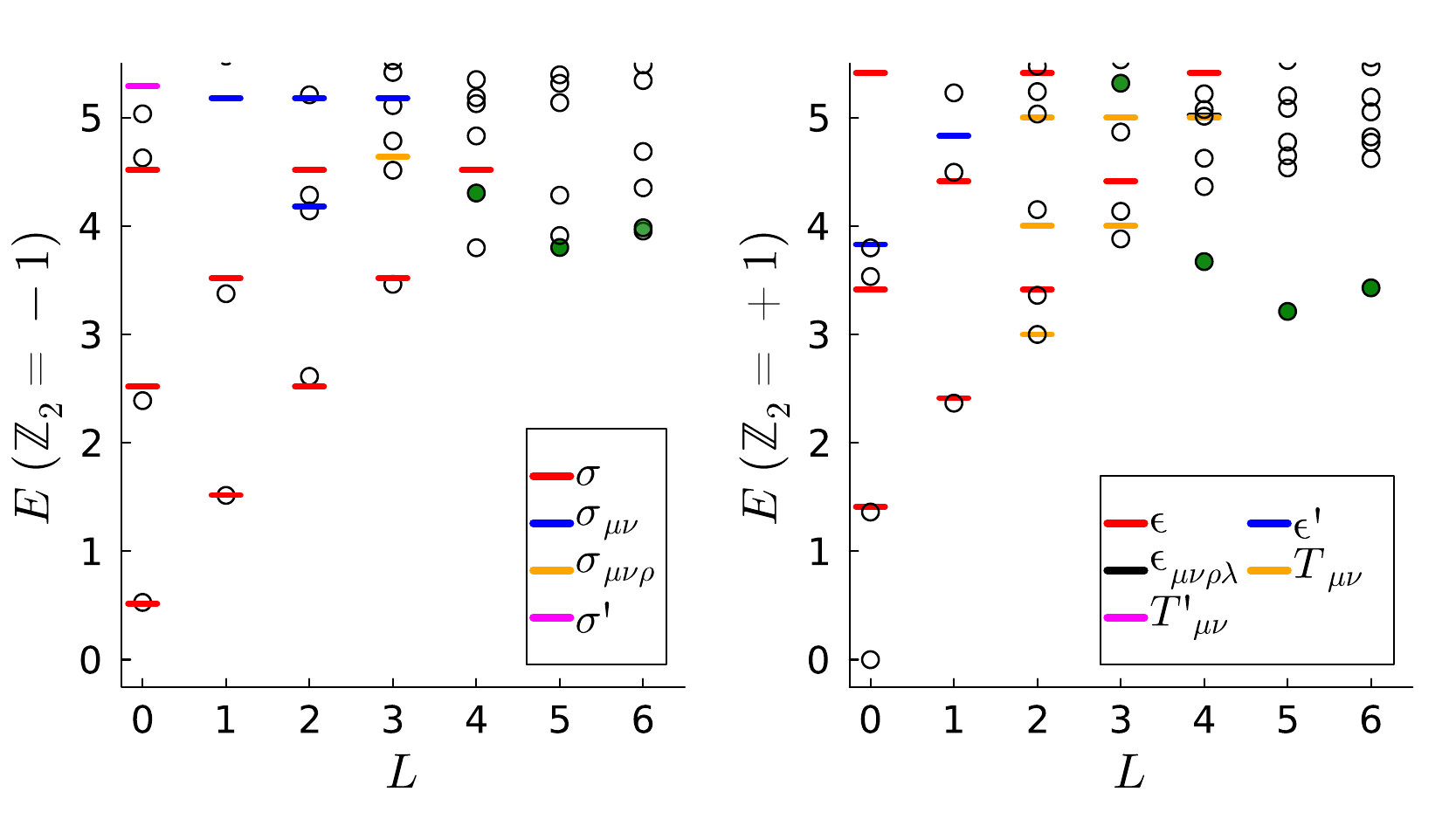}
    \caption{Spectrum of the model in Eq.~\eqref{eq:interaction} at  $h=0.183$ where the coupling to the $\epsilon$ perturbation vanishes at the given system size, $N=8$. The ED data (empty circles) are rescaled such that $E_T = 3$. Filled green circles are identified by having high overlap with the MF magnetoroton states ($>0.85$, with the minimum attained at $L=3$). For low energies and spin, we observe a one-to-one correspondence between states and CFT operators, as expected, and their energies are expected to agree as $N\to\infty$.}
    \label{fig:conformal_spectrum_hc_Q}
\end{figure}

After obtaining some insight into the effect of conformal perturbations, we make another attempt at extracting the state-operator correspondence.  In \cref{fig:conformal_spectrum_hc_Q} we present the spectrum for $N=8$ particles at its own critical field value $h_c(Q) \approx 0.183$, which was determined in \cref{fig:coupling_epsilon}. At low energies and angular momenta, we notice a one-to-one correspondence between the microscopic states and the Ising CFT operators, now with better agreement with their scaling dimensions. Similar to Fig.~\ref{fig:conformal_spectrum_hc_inf}, in both parity sectors we still observe the non-CFT magnetoroton states that can be clearly identified by their overlaps with the MF states, even when they are close in energy to CFT states (e.g., at $L=3$). 

\subsection{The drift of the critical point}

One immediate question concerns the drift in the values of $h_c(Q)$ extracted in \cref{fig:coupling_epsilon}, and whether further fine-tuning can remove it. One cause for the drift could be the coupling between the spin and the neutral excitations of the $\nu=1/3$ state that are absent at $\nu=1$. To test this hypothesis, we make use of the fact that the energy of the magnetoroton can be tuned by varying $V_1^\text{intra}$ pseudopotential. In Appendix~\ref{app:roton} we repeat the conformal perturbation analysis for different values of $V_1^\text{intra}$, finding that the drift in Fig.~\ref{fig:coupling_epsilon} remains present even when the magnetoroton is completely gapped out from the low-energy spectrum. This suggests that a more likely explanation for the drift is the significant difference in the correlation length, when compared to the IQH case. A particular issue sensitive to this correlation
length was discussed in the recent study \cite{Lauchli25}, where similar drifts in models at $\nu=1$ with different values of $V_0^\text{inter}$ were attributed to the curvature of the fuzzy sphere. This points towards a generic feature that is not specific to our anyonic model. Surprisingly, tuning the drift to zero at $\nu=1$ coincides with an almost vanishing $g_{\epsilon'}$, where the conformal spectrum can be matched with very high accuracy. We have found this to not be the case in the $\nu=1/3$ model, as removing the drift in $g_\epsilon$ significantly increases the value of $g_{\epsilon'}$ and leads to large deviations in the spectrum.

%The consistency of the MF approach even at the critical point provides additional evidence that the charge sector is effectively decoupled from the spin sector. Nevertheless, one might wonder if the residual coupling between the two sectors in finite sizes might be responsible for the drift of $h_c(Q)$ extracted from the $\epsilon$ perturbation in Fig.~\ref{fig:coupling_epsilon}. To test this hypothesis, we make use of the fact that the energy of the magnetoroton can be tuned by varying $V_1^\text{intra}$ pseudopotential. In Appendix~\ref{app:roton} we repeat the conformal perturbation analysis for different values of $V_1^\text{intra}$, finding that the drift in Fig.~\ref{fig:coupling_epsilon} remains present even when the magnetoroton is completely gapped out from the low-energy spectrum. This suggests that a more likely explanation for the visible drift of the critical point in Fig.~\ref{fig:coupling_epsilon}, compared to the IQH case, is the significant difference in the correlation length in the two cases. This difference, combined with the smaller range of system sizes accessible in the FQH case, implies that the finite-size effects in capturing the vacuum state are a  predominant cause of the drift in Fig.~\ref{fig:coupling_epsilon}. 

Another interesting feature in \cref{fig:coupling_epsilon} is the approximate collapse of the finite-size $g_\epsilon$ curves around the value $h_c \approx 0.12$. The finite-size crossings of the order parameter curves around the same point seen in \cref{fig:order_parameter} is not coincidental, and the two can be linked using conformal perturbation. 
At first order, the conformal vacuum mixes with other even parity states: 
\begin{equation}
    |0\rangle \propto |\mathbb{I}\rangle - \sum_{\mathcal{V}}\frac{g_\mathcal{V}}{\Delta_\mathcal{V}}|\mathcal{V}\rangle + \ldots \, ,
\end{equation}
where the sum includes only scalar primaries, and the ellipsis stand for higher-order contributions from the descendants. Accordingly, the value of the order parameter changes as:
\begin{equation}
    \frac{\langle M^2 \rangle}{R^{4-2\Delta_{\sigma}}} \propto \langle \mathbb{I}| \sigma \sigma |\mathbb{I}\rangle - 2\sum_{\mathcal{V}}\frac{g_\mathcal{V}}{\Delta_\mathcal{V}} \langle \mathbb{I} | \sigma \sigma |\mathcal{V}\rangle + \ldots \, ,
\end{equation}
where the omitted proportionality constant has no system-size dependence, and the symbols $\langle \mathbb{I}| \sigma \sigma |\mathbb{I}\rangle$ and $\langle \mathbb{I}| \sigma \sigma |\mathcal{V}\rangle$ represent quantities that only depend on the scaling dimension of $\Delta_\sigma$ and $\Delta_\mathcal{V}$.
The corrections are thus proportional to the OPE coefficients $f_{\sigma\sigma\mathcal{V}}$, with $f_{\sigma\sigma\epsilon} \approx 1.051853$ being by far the largest one involved. Consequently, the crossing in the rescaled order parameter can be attributed to a collapse in the finite-size values of $g_\epsilon$, even though this appears at a non-zero value. 

\section{Entanglement Entropy and the  F-Theorem}\label{sec:f-theorem}

Our results so far hint towards a ``clean'' separation between the charge and spin degrees of freedom: the FQH spectrum does not strongly perturb the CFT spectrum, even when the two coexist at similar energies. In this section, we probe this further using bipartite entanglement entropy
$S_A = - \mathrm{tr} \rho_A \ln \rho_A$,
where $\rho_A=\mathrm{tr}_{\bar A}|\psi\rangle\langle \psi|$ is the reduced density matrix of the subsystem $A$, obtained by tracing its complement $\bar A$. We choose a real space bipartition with  $A$ being a spherical cap defined by the polar angle $\theta_A$~\cite{Sterdyniak12,Dubail12a}. 

In 2+1D, both gapped and gapless (described by a CFT) systems obey the area law for bipartite entanglement entropy:
\begin{eqnarray}
 S_A    = \eta R \sin \theta_A - \gamma \, ,
\end{eqnarray}
where the proportionality constant $\eta$ is model dependent and $\gamma$ is a universal subleading constant term that can contain information about both spin and charge sectors. For states described by CFT, $\gamma$ is known as the F-function and it decreases monotonically along any RG flow~\cite{Jafferis2011,Myers2011,Casini2011}. It can be viewed as a higher-dimensional generalization of a renormalization group irreversibility known as the c-theorem in 2D CFTs~\cite{ZamolodchikovCtheorem,Casini07}. On the other hand, in generic gapped states, the F-function is also known as the topological entanglement entropy~\cite{KitaevPreskill,LevinWen}. 

The 3D Ising F-function is less than the value of $\gamma$ for any topologically ordered state~\cite{Grover14}. Hence, the transition out of any topologically ordered phase cannot be simply captured by a single Ising CFT. Consequently, the topological entanglement entropy of any underlying FQH state in the charge sector must remain constant throughout the transition, and any changes can be attributed to the spin degree of freedom.  In the fuzzy sphere bilayer model, Eq.~\eqref{eq:interaction}, the evolution of the universal constant can be understood as follows. The Laughlin paramagnet has $\gamma_\text{para} = \gamma_\text{topo} = \ln \sqrt{3}$, while in the $\mathbb{Z}_2$ Laughlin ferromagnet we expect $\gamma_\text{ferro} = \gamma_\text{topo} - \ln 2$. The $\ln 2$ correction comes from working in the even parity sector such that $|\psi\rangle = (|\psi_{M=N}\rangle + |\psi_{M=-N}\rangle)/\sqrt{2}$ and ensures there is no discontinuity in $\gamma$. At the critical point, we  therefore expect $\gamma_\text{critical} = \gamma_\text{topo} + F_\text{Ising}$. 

We implement the method of Ref.~\cite{Hu24a} for the extraction of the universal constant: 
\begin{equation}
    \gamma = (\tan \theta \partial_\theta - 1) S_{A}|_{\theta = \pi/2} \, .
\end{equation}
Similarly to the $\nu=1$ model, to observe the RG flow of the F-function in the Ising CFT, we need to account for the charge sector by subtracting the single layer entanglement entropy. In the integer case, the correction denoted by $\gamma_\text{IQH}(Q)$ can be calculated efficiently \cite{Rodriguez09} and vanishes in the thermodynamic limit, $\gamma_\text{IQH}(Q \to \infty) = 0$. However, in our case the correction needs to be field-dependent -- the effective single-layer interaction depends on the polarization (due to the non-zero inter-layer $V_3$ pseudopotential), and this affects how the topological entropy of the Laughlin state approaches the thermodynamic value of $\gamma_\mathrm{topo}$. We therefore resort to the MF approximation of \cref{sec:mean-field}, using the predetermined optimal polarizations $\theta_\text{opt}$, to compute the corresponding corrections $\gamma_\text{FQH}(Q)$.

\begin{figure}[t]
    \centering
    \includegraphics[width=0.48\textwidth]{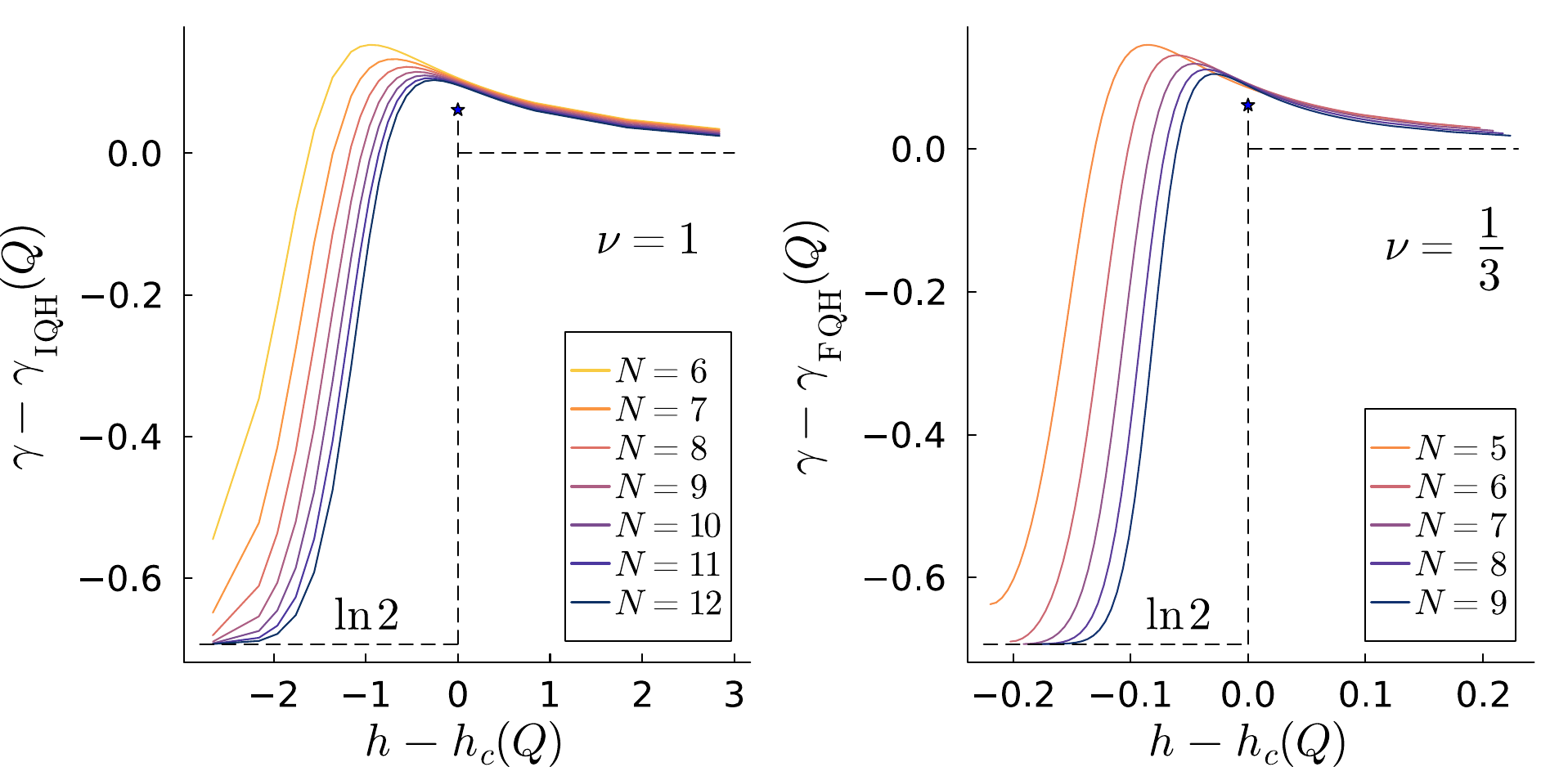}
    \caption{Constant contribution to the regularized entanglement entropy in the bilayer model, Eq.~\eqref{eq:interaction}. For $h<h_c$, the constant is expected to be $\ln 2$, as the ground state approaches a macroscopic superposition of fully polarized states. For $h=h_c$, the contribution equals the F-function of the underlying CFT, approximately equal to $F_\mathrm{Ising} = 0.0612$ (the blue star). At $h > h_c$, the F-function approaches zero.
    Left panel: The $\nu=1$ model. The subtracted entropy $\gamma_\text{IQH}(Q)$ corresponds to a single-layer IQH state, and is independent of the field $h$. For the critical field we use $h_c(Q) = 3.16$ for all $Q$, as there appears to be no significant drift~\cite{Zhu23}.
    Right panel: The $\nu=1/3$ model. The correction $\gamma_\text{FQH}(Q)$ corresponds to a single-layer FQH state, and is now dependent on the field, according to the optimal polarization $\theta_\text{opt}$. The values of the critical field are those extracted in \cref{fig:coupling_epsilon} with conformal perturbation theory.}
    \label{fig:f-function}
\end{figure}

A direct comparison between $\nu=1$ and $\nu=1/3$ is shown in \cref{fig:f-function}. After applying the regularization scheme, we observe nearly identical behavior of the universal constant in the two models, reinforcing the idea that, in this case, the charge sector is ``invisible'' to the Ising CFT. This serves as a numerical demonstration of the F-theorem in 2+1D. A non-perturbative estimation of the Ising F-function can be obtained by extrapolating the value of $\gamma$ at the critical point; this was carried out at $\nu=1$, obtaining $F_\mathrm{Ising} = 0.0612(5)$ \cite{Hu24a}, very close to that obtained through the $4-\epsilon$ expansion \cite{Fei15}. Unfortunately, at $\nu=1/3$ the accessible system sizes preclude a reliable extrapolation of the F-function.

\section{Conclusions}

We have presented evidence for the 3D Ising transition in a fractionally-filled fuzzy sphere model, where the charge sector realizes a strongly-correlated state with topological order. We have illustrated our approach with the $\nu=1/3$ Laughlin state, while the Appendices show that similar results are obtained for both fermionic and bosonic FQH states with different kinds of topological order. We have applied conformal perturbation theory to extract CFT data with improved accuracy. The real-space entanglement entropy was used to show the decoupling of the charge and spin sectors, illustrating the robustness of the fuzzy sphere regularization.

The approach presented here opens up a number of interesting directions. An immediate question concerns the possibility of realizing other kinds of CFTs beyond the 3D Ising using FQH states as a platform. While our approach is expected to work more generally, the formulation of the effective interactions that give rise to suitable FQH states is subtle and needs to be verified on a case-by-case basis. Furthermore, given that the charge and spin degrees of freedom can be separated efficiently, could models be engineered where the interactions of the two lead to new CFTs? The recent study~\cite{Zhou24b} has revealed possible parity-breaking CFTs by fractionally filling multiple flavors of particles. Our approach also lays the foundation for identifying conformal critical points in various models that are relevant for FQH bilayer experiments~\cite{Read00,Barkeshli2010,Barkeshli2011,Zhang2023XYstar,Wu2023XYstar}, where the interactions can be conveniently tuned by changing the distance between FQH layers~\cite{Eisenstein14}, their widths~\cite{Singh2024}, imbalance of charge~\cite{Li2019} and even the underlying band structure via doping~\cite{Wang2024Holes}. Finally, an interesting question is whether stronger coupling between the CFT and FQH spectra could be induced. Such a situation could arise if the charge sector is described by a FQH nematic state~\cite{Mulligan10,Maciejko13,You14,Pu2024Nematic}, which is gapped to charge excitations, while it also has a charge-neutral Goldstone mode due to a spontaneously broken continuous rotational symmetry. 

\section*{Acknowledgements}

We would like to thank Ajit C. Balram, Junkai Dong, Benoit Estienne, Johannes Hofmann, Wei Zhu and Ashvin Vishwanath for enlightening discussions on related topics. C.V. and Z.P. acknowledge support by the Leverhulme Trust Research Leadership Award RL-2019-015 and EPSRC Grant EP/Z533634/1. 
Statement of compliance with EPSRC policy framework on research data: This publication is theoretical work that does not require supporting research data. 
This research was supported in part by grant NSF PHY-2309135 to the Kavli Institute for Theoretical Physics (KITP). Z.P. acknowledges support by the Erwin
Schrödinger International Institute for Mathematics and Physics.
Computational portions of this research have made use of DiagHam~\cite{diagham} and FuzzifiED~\cite{FuzzifiED} software libraries, and they were carried out on ARC3 and ARC4, part of the High-Performance Computing facilities at the University of Leeds. The Flatiron Institute is a division of the Simons Foundation. R.F. is supported by the Gordon and Betty Moore Foundation (Grant GBMF8688).

\appendix

\section{Model optimization}\label{app:optimization}

The difficulty of locating the optimal point for observing the conformal transition is considerably increased once more interaction pseudopotentials are introduced in the model. Fortunately, we are able to restrict the space of pseudopotentials directly relevant to the transition to a small subset by fixing the rest. In this Appendix, we exemplify this method for the $\nu=1/3$ model, while generalizations to other fractions are presented in the subsequent Appendix~\ref{app:othernu}. 

We begin by fixing the overall energy scale by setting $V_1^\text{intra} = V_1^\text{inter}=1$. Furthermore, for fermions we can neglect the even intra-layer pseudopotentials, hence we can set $V_0^\text{intra} = 0$. Although higher odd psuedopotentials do have an effect on the spectrum, they are not necessary, thus we set them to zero. In order to have a gapped state in $h=0$ limit, we require a non-zero  $V_0^\text{inter}$. Its precise value was found to have  a weak effect on the spectrum at the critical point and we set $V_0^\text{inter}=1$ (for example, the low-energy spectrum is qualitatively unchanged for $V_0^\text{inter}=10$ even near $h_c$). This leaves the question: how far in inter-layer pseudopotential range do we need to go? The minimal model must contain $V_3^\text{inter}$, as this ensures the Ising $\mathbb{Z}_2$ symmetry is not accidentally enlarged. To see this, we look at the interaction matrix elements~\cite{Fano86}: 
\begin{equation}
\begin{aligned}
    V_{j_1 j_2 j_3 j_4} &= \sum_l V_m (4Q-2m+1) \delta_{j_1 + j_2, j_3+j_4} \times \\ & \begin{pmatrix}
        Q & Q & 2Q-m \\ j_1 & j_2 & -j_1 - j_2
    \end{pmatrix}
    \begin{pmatrix}
        Q & Q & 2Q-m \\ j_3 & j_4 & -j_3 - j_4
    \end{pmatrix} \, .
\end{aligned}
\end{equation}
The symmetry property of the Wigner $3j$ symbols, in combination with the fermionic statistics, allow us to split the inter-layer interaction into singlet (for even pseudopotentials) and triplet actions (for odd pseudopotentials):
\begin{align}
    H_\text{inter} &= \sum  V_{j_1 j_2 j_3 j_4}^\text{odd} (c^\dagger_{j_1 \uparrow}c^\dagger_{j_2 \downarrow} + c^\dagger_{j_1 \downarrow}c^\dagger_{j_2 \uparrow}) (c_{j_3\uparrow}c_{j_4 \downarrow} + c_{j_3 \downarrow}c_{j_4 \uparrow}) \nonumber \\
    &+  V_{j_1 j_2 j_3 j_4}^\text{even} (c^\dagger_{j_1 \uparrow}c^\dagger_{j_2 \downarrow} - c^\dagger_{j_1 \downarrow}c^\dagger_{j_2 \uparrow}) (c_{j_3\uparrow}c_{j_4 \downarrow} - c_{j_3 \downarrow}c_{j_4 \uparrow})
\end{align}
From this, it becomes evident that the minimal model at $\nu=1/3$ needs a $V_3^\text{inter}$ term. 
Otherwise, the ground state takes the form of a degenerate $S=N/2$ Laughlin multiplet, regardless of the ratio $V_1^\text{intra}/V_1^\text{inter}$.

\begin{table}[t]
    \centering
    \begin{tabular}{|m{7.5em}||m{3em}|m{3em}|m{3em}|m{3em}|m{3em}|m{3em}|}
    \hline
         & $\sigma$ & $\epsilon$ &  $\epsilon'$ & $\sigma_{\mu \nu}$ & $\sigma_{\mu\nu\rho}$  \\
        \hline
        \flushleft Fuzzy Sphere & 0.528 & 1.362 & 3.798 & 4.139 & 4.514\\
        \hline
        \flushleft Bootstrap & 0.518 & 1.413 & 3.830 & 4.180 & 4.638\\
        \hline
        \flushleft Relative error & 1.9\% & 3.6\% & 0.8\% & 1.0\% & 2.7\%\\
        \hline
    \end{tabular}
    \caption{Scaling dimension comparison between conformal bootstrap and the $\nu=1/3$ fuzzy sphere model with $N=8$ particles. The lowest 5 primaries are included, except the stress-energy tensor $T_{\mu\nu}$.}
    \label{tab:scaling_dimensions}
\end{table}

In summary, we are left with three parameters to optimize over: $V_2^\text{inter}, V_3^\text{inter}$ and $h$. To find the optimal tower structure, we perform a simple gradient descent, where the cost function is the sum of squared differences between ED energies and conformal bootstrap: 
\begin{equation}\label{eq:cost_fn}
    \delta = \sum{(E_i - \Delta_i)^2} \, ,
\end{equation}
where the energies $E_i$ have been rescaled such that $E_T=3$, and the set of states we optimize over is $\{\sigma, \partial \sigma, \partial \partial \sigma, \square \sigma\}$ in the odd parity sector, and $\{\epsilon, \partial \epsilon, \partial \partial \epsilon, \square \epsilon\}$ in the even parity sector.

The optimal point at system size $N=8$ was found to be $V_3^\mathrm{inter} \approx 0.087$, $V_2^\mathrm{inter} \approx 0.488$, $h\approx 0.185$. The optimal values of $V_2^\mathrm{inter}, V_3^\mathrm{inter}$ slightly drift with system size; to compare with bootstrap data, we rely on conformal perturbation, as detailed in the main text. For the optimal value of the $h$ field predicted by the $\epsilon$ perturbation in system size $N=8$ shown in Fig.~\ref{fig:conformal_spectrum_hc_Q}, we list the scaling dimensions of the lowest five primaries in Table~\ref{tab:scaling_dimensions}.

\section{Charge and spin gaps}\label{app:gaps}

A prerequisite for observing the 3D Ising universality class is a non-zero charge excitation gap. In a state with fractionalized excitations such as the Laughlin state, the charge gap is defined as the energy cost of creating a quasiparticle (at flux $2Q-1$) and a quasihole (at flux $2Q+1$). The charge gap is defined as 
\begin{equation}
\begin{aligned} 
    \Delta E_c &= \tilde{E}_0(2Q-1) + \tilde{E}_0(2Q+1) - 2\tilde{E}_0(2Q) \, ,\\
    \tilde{E}_0(2Q) &= E_0(2Q) - C_{2Q}(N^2 - n_q^2 e_q^2)/2 \, ,
\end{aligned}
\label{eq:charge gap}
\end{equation}
where $E_0(2Q)$ is the ground state energy at a given flux and $\tilde{E}_0(2Q)$ is the total energy that includes a uniform neutralizing background charge~\cite{Morf02, Jain97}.   The correction factor in the total energy depends on the number of charge excitations ($n_q = 1$ in the quasihole/quasiparticle case and $n_q=0$ in the vacuum), on their charge (e.g., $e_q = 1/3$ for the Laughlin state with $m=3$), and on the average charging energy per particle pair \cite{Balram20b}:
\begin{equation}
    C_{2Q} = \sum_{m=0}^{2Q} V_m \frac{4Q - 2m +1}{(2Q+1)^2} \,. 
\end{equation}
For a generic value of the transverse field $h$, our model does not conserve the layer particle number, therefore we use the averaged pseudopotentials $V_m = (V^\text{intra}_m + V^\text{inter}_m)/2$. 

\begin{figure}
    \centering
    \includegraphics[width=0.48\textwidth]{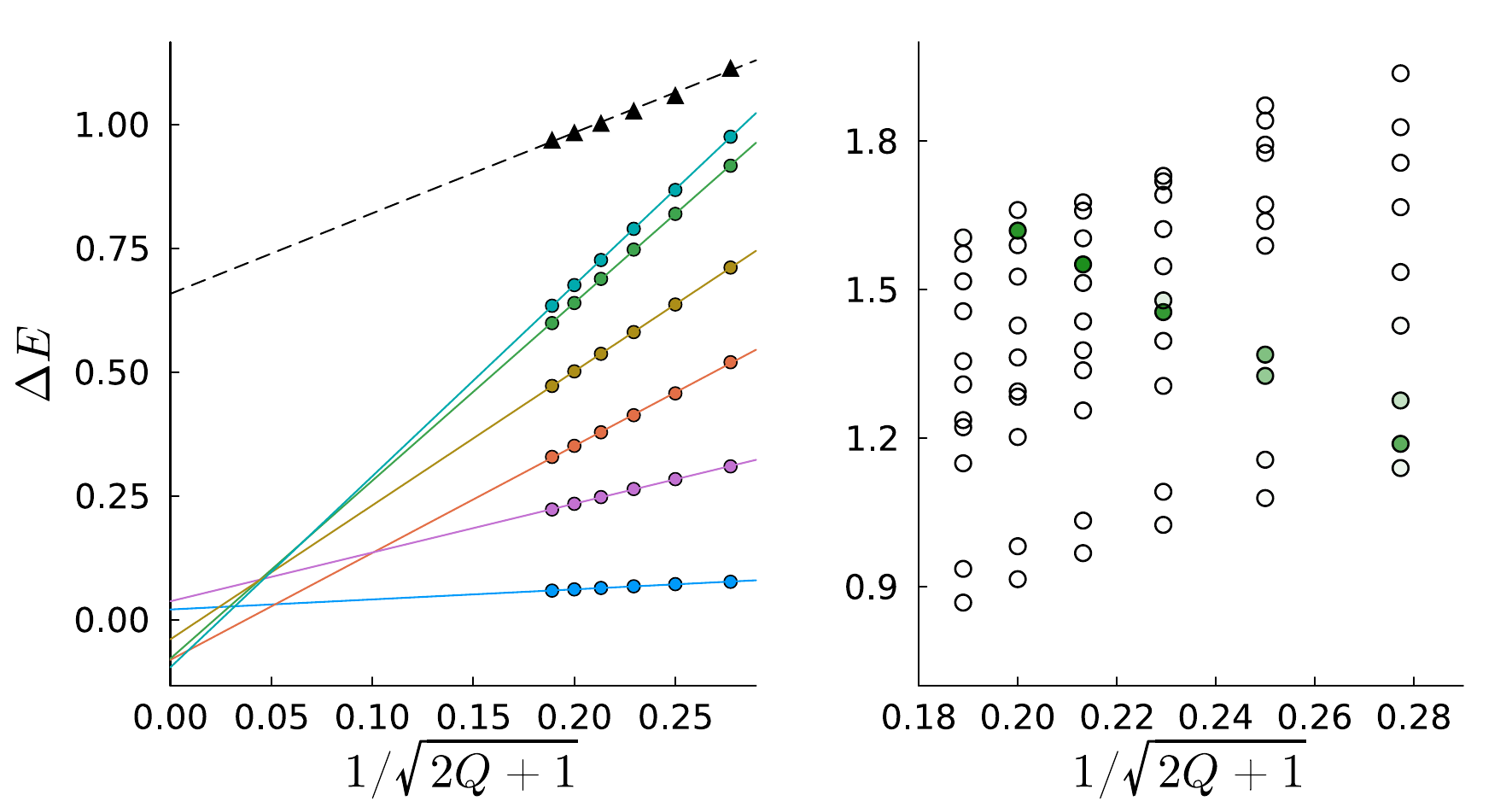}
    \caption{(a) Finite size scaling of the charge and spin gaps at the phase transition point $h_c = 0.135$, for system sizes $N=5-10$. The charge gap (black triangles), Eq.~(\ref{eq:charge gap}), converges to a finite value in the thermodynamic limit. The spin gaps in each of $L \in \{0,1,2 \}$ sectors with either $\mathbb{Z}_2$-parity (colored circles) vanish in the thermodynamic limit. We apply the density correction, $\sqrt{2Q \nu /N}$, to all gaps, which accounts for the deviation of the particle density in a finite system compared to the thermodynamic limit~\cite{Morf86}. (b) The neutral excitation gaps in $L=3$, $\mathbb{Z}_2$-even sector. Most states belonging to the CFT sector have polynomially decreasing gaps (empty circles), while the roton state (green circle) has energy that increases with system size. }
    \label{fig:gaps_scaling}
\end{figure}

The computed charge gaps for different system sizes are presented in \cref{fig:gaps_scaling}(a) and show good convergence to a finite value in the thermodynamic limit. On the other hand, the gaps of CFT states in the neutral sector are expected to vanish in the thermodynamic limit. \cref{fig:gaps_scaling}(a) contrasts the finite-size scaling of the first excited states in $L\in \{ 0,1,2\}$ sectors with the charge gap defined in Eq.~(\ref{eq:charge gap}). All of the neutral gaps extrapolate to a value much smaller than the charge gap and in the vicinity of zero. Some of the extrapolated values are slightly below zero, which is attributed to the uncertainty of the extrapolation due to limited system sizes available.

At higher angular momenta, we expect that the magnetoroton states will appear at lower energies and mix  with the CFT spectrum. In \cref{fig:gaps_scaling}(b), we show the energies of all low-lying states with $L=3$ in the $\mathbb{Z}_2$-even sector. In the smallest system size, the roton state -- identified by high overlap with the MF magnetoroton state -- appears as the second-lowest state in the given angular-momentum spectrum. However, the trend of its energy dependence on system size is opposite to that of CFT states: while the latter decrease with $N$, the roton energy grows with $N$.

%\begin{figure}
%    \centering
    %\includegraphics[width=0.48\textwidth]{Figs/prx_spin_gaps_roton.pdf}
    %\caption{Finite size scaling of the neutral gaps in the $L=3$, $\mathbb{Z}_2$-even sector for system sizes $N=5-10$, calculated at the finite-size critical fields $h_c(Q)$. The magnetoroton state, identified with the mean-field treatment of \cref{sec:mean-field}, is seen to increase in energy while the other states (part of the CFT sector) all have closing gaps in the thermodynamic limit. }
    %\label{fig:spin gaps roton}
%\end{figure}

\section{Interplay between charge and spin degrees of freedom}\label{app:roton}

In this Appendix, we examine the effect of the charge sector on degrees of freedom in the spin sector. In particular, we inquire whether the low-lying gapped excitations of the Laughlin state, i.e., the magnetoroton, could affect the CFT spectrum, and even cause the systematic drift in $h_c(Q)$ that was absent in the original study at unit filling~\cite{Zhu23}. 

\begin{figure}[tb]
    \centering
    \includegraphics[width=0.48\textwidth]{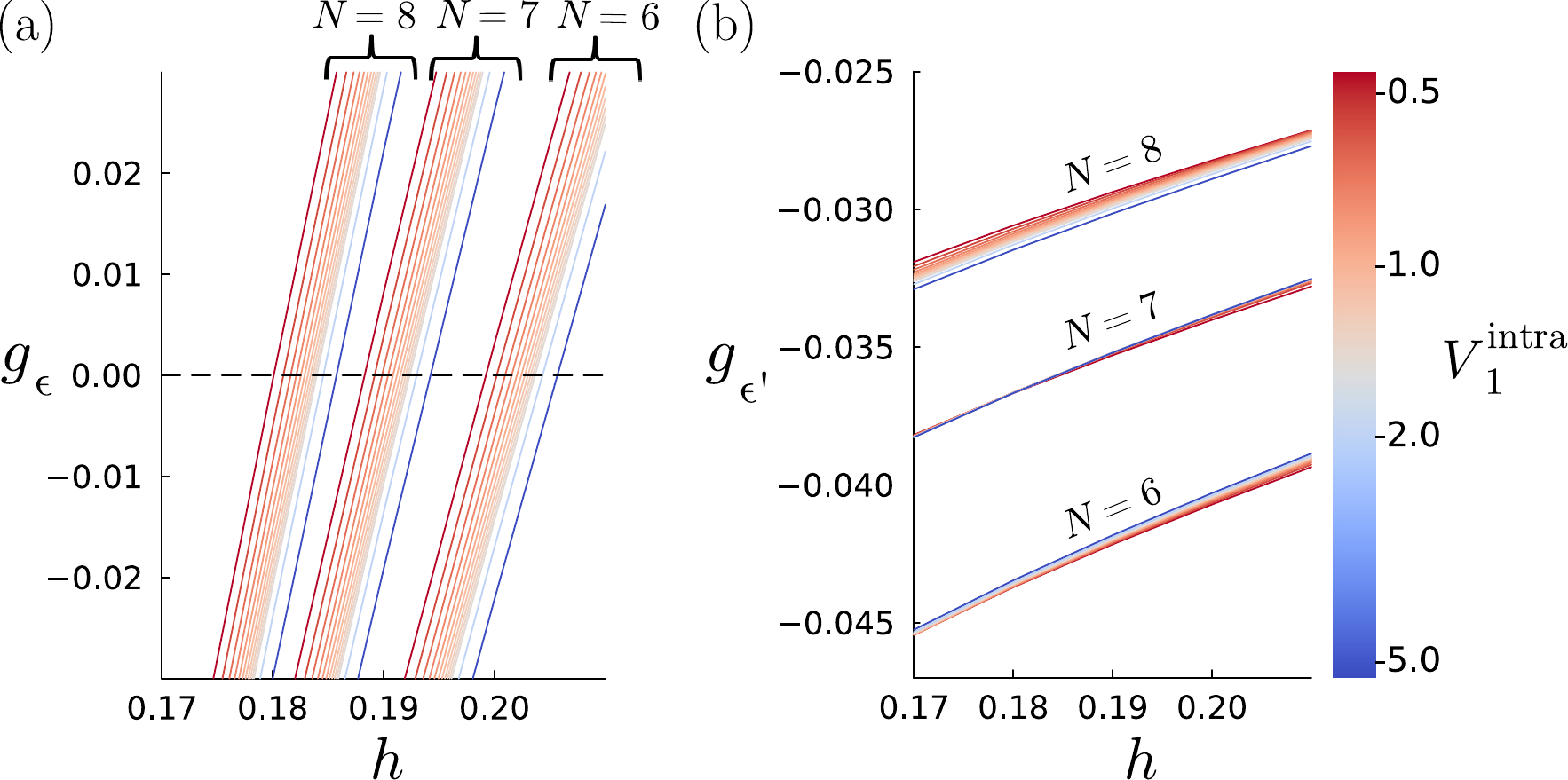}
    \caption{Conformal perturbation couplings, $g_\epsilon$ in (a) and $g_{\epsilon'}$ in (b), as a function of $h$ for different values of $V_1^\text{intra} \in [0.5,5]$. The effect of varying $V_1^\text{intra}$ on the couplings is small and inversely proportional to the roton gap. Moreover, even as the magnetoroton gap is tuned to energies much higher than the energies of the low-lying CFT states, the $h_c(Q)$ still shows significant drift. We note that lowering $V_1^\mathrm{intra}$ below 0.5 makes the underlying Laughlin state unstable. }
    \label{fig:v1_intra_tune}
\end{figure}

As shown in \cref{sec:spectrum}, we can clearly identify magnetoroton excitations on top of both the vacuum and the $\sigma$ state even in the presence of CFT states at similar energies.
This observation motivates the following assumption. Near the critical point, the low-energy part of the Hilbert space is a tensor product of the CFT part and a gapped part coming from the FQH degrees of freedom, $\mathcal{H} \approx \mathcal{H}_\text{CFT} \otimes \mathcal{H}_\text{gapped}$. 
Assuming a perturbative coupling between the two sectors, the low-energy effective Hamiltonian reads
\begin{equation}\label{eq:rotonPT}
\begin{gathered}
    H = H_0 + V \\
    H_0 = H_\text{CFT} + H_\text{gapped}\,,\,
    V = \lambda \int \mathrm{d}^2\Omega \, \epsilon(\Omega) \delta\rho(\Omega),
\end{gathered}
\end{equation}
where $\lambda$ is the coupling strength, $\epsilon$ the $\mathbb{Z}_2$-even primary field in the CFT, $\delta\rho$ an operator that creates the magnetoroton excitation. 
Under the SMA, which is known to be valid over only a limited range of momenta smaller than the roton minimum~\cite{Yang12b}, $\delta\rho(\Omega)$ is the density fluctuation operator as we have analyzed in \cref{eq:sma}.
Noting that the rotation in the CFT descends from the magnetic translation in the LLL, one can verify that $V$ is indeed the simplest symmetry-allowed coupling between the two sectors.
%which can be constructed from \cref{eq:sma}. Eq.~(\ref{eq:rotonPT}) is an approximation and assumes that SMA holds, which is known to be valid over only a limited range of momenta smaller than the roton minimum~\cite{Yang12b}. 

Upon projection into the CFT subspace, the perturbation in Eq.~(\ref{eq:rotonPT}) may generate additional $\epsilon$ terms that affect the finite-size critical fields $h_c(Q)$; this can be shown by integrating out the magnetoroton modes via a standard second order perturbation approach.
Let $|\alpha\rangle$ denote the CFT states and $|\psi_{1/q}\rangle$ the FQH vacuum. Then, we introduce the spectrum projector onto the low-energy space $P = \sum_\alpha |\alpha \rangle\langle \alpha| \otimes |\psi_{1/q} \rangle\langle \psi_{1/q}|$, and its complement $Q = 1 - P$. The effective Hamiltonian reads
\begin{equation}
    H_\text{eff} = PH_\text{0}P + PVQ\frac{1}{E - QH_0Q}QVP \, ,
\end{equation}
%the magnitude of these terms will be inversely proportional to the magnetoroton gap, $\delta g_\epsilon \sim 1/ E^\text{roton}_{L}$. }
where the perturbed energy $E$ is determined self-consistently through $H_\text{eff}|\psi\rangle = E|\psi\rangle$. In the regime where the energy of exciting magnetoroton is large, it suffices for a qualitative analysis to assume a flat dispersion for the whole branch and have $E - QH_0Q \approx -\Delta_{\text{MR}}$. 
We write the perturbation more explicitly as 
\begin{equation}
    PVQVP = \lambda^2 \int \varepsilon(\Omega_1)\varepsilon(\Omega_2) G^{(q)}(\Omega_{12}) \,\mathrm{d}\Omega_1\mathrm{d}\Omega_2 \, ,
    \label{eq:PVQVP}
\end{equation}
with $G^{(q)}(\Omega_{12}) = \langle \psi_{1/q}|\delta\rho(\Omega_1)\delta\rho(\Omega_2)|\psi_{1/q}\rangle$ being the two-point function of the density fluctuation. The gap of the magnetoroton mode enforces an exponential decay for $G^{(q)}(\Omega_{12})$ at long distances \cite{Girvin86}, and the right-hand side of \eqref{eq:PVQVP} is dominated by the short-distance contributions. 
Accordingly, we can replace the product of the CFT primaries by their OPE
\begin{equation}
\epsilon(\Omega_1)\epsilon(\Omega_2) = f_{\epsilon\epsilon\epsilon} |\Omega_{12}|^{-\Delta_\epsilon}\epsilon(\Omega_C) + \ldots
\end{equation}
where $\Omega_C$ is the center-of-mass coordinate, and replace $G^{(q)}$ by its short-distance form
\begin{equation}
G^{(q)}(\Omega_{12}) \approx \rho_{1/q}^2(-1 + A_{q} |\Omega_{12}|^{2q})\,,\, A_q > 0 \,.
\label{eq:Gq short distance}
\end{equation}
%\zp{Is the above for any filling $1/q$ or just 1/3? (the formulas contain a mix, please check if it's all fine)} 
%On the CFT side, the generation of new terms can be seen by applying the OPE: $\epsilon(\Omega_1)\epsilon(\Omega_2) = f_{\epsilon\epsilon\epsilon} |\Omega_{12}|^{-\Delta_\epsilon}\epsilon(\Omega_C) + \dots$, where $\Omega_C$ is the center-of-mass coordinate. In terms of the magnetoroton gap $\Delta_\text{MR}$, we find
Namely, we have
\begin{equation*}
    H_\text{eff} \approx H_\text{CFT} - \lambda^2 \frac{f_{\epsilon\epsilon\epsilon}}{\Delta_\text{MR}} \int \epsilon(\Omega_C) \mathrm{d}\Omega_C\int \frac{G^{(q)}(\Omega_{12})}{|\Omega_{12}|^{\Delta_\epsilon}}\mathrm{d}\Omega_{12} 
\end{equation*}
where $G^{(q)}(\Omega_{12})$ takes the short-distance form \eqref{eq:Gq short distance}.
Noting that $G^{(q)} < 0$, the integral over $\Omega_{12}$ is also negative. We have
\begin{equation}
    H_\text{eff} \approx H_\text{CFT} + \lambda^2 f_{\epsilon\epsilon\epsilon} \frac{a \rho_{1/q}^2}{\Delta_\text{MR}}\int \epsilon(\Omega) \mathrm{d}\Omega \, ,
    \label{eq:rotonPTscaling}
\end{equation}
where $a$ is a {\em positive} constant arising from the second integral, whose specific value is less important.

As expected, the coupling between the CFT and the gapped neutral modes of the FQH sector generates an additional contribution to the effective coupling $g_\epsilon$, which, consequently, shifts the critical point $h_c(Q)$ by a constant amount.
Notably, such a generated perturbation \emph{decreases} inversely proportional to the magnetoroton gap.
In our microscopic model we test this by varying the roton gap through the pseudopotential $V_1^\text{intra}$ and the results are shown in \cref{fig:v1_intra_tune}. The $g_\epsilon$ coupling decreases as the strength of $V_1^\text{intra}$ is increased, in agreement with the results of \cref{eq:rotonPTscaling} and thus confirming the treatment of \cref{eq:rotonPT}. However, as we see from both the analytical and numerical analysis, the change is constant across system sizes, meaning that the coupling to the magnetoroton does not account for the drift in $h_c(Q)$. 

A more likely explanation of the drift in $h_c(Q)$ may lie in the structure of the vacuum, rather than the excitations of the charge sector. Indeed, there is a significant difference in correlation length between the IQH and FQH states, e.g., $\xi\approx 1.4\ell_B$ in the Laughlin $\nu=1/3$ state \cite{Estienne15}; thus, we might expect that the restriction to smaller system sizes compared to $\nu=1$ causes a stronger effect on $h_c(Q)$. % A particular issue sensitive to this correlation length was discussed in the recent study \cite{Lauchli25}, where the drifts in models at $\nu=1$ with different values of $V_0^\text{inter}$ were attributed to the curvature of the fuzzy sphere. This suggests that the drift is a general feature that is not specific to our anyonic model. Surprisingly, tuning the drift to zero at $\nu=1$ coincides with an almost vanishing $g_{\epsilon'}$, which does not appear to be the case in our FQH models.

\section{Different filling factors}\label{app:othernu}

In this Appendix we briefly investigate the existence of the 3D Ising transition, discussed in the main text, for a few other filling factors and particles with bosonic exchange statistics. We will show that the 3D Ising CFT can be successfully encoded onto different charge ``substrates'', either by placing electrons at different fractional fillings or by taking bosonic particles instead of electrons. We recall that bosonic FQH states can be generally defined by taking a fermionic FQH wave function and dividing through by an overall Jastrow factor, $\prod_{i<j}(u_i v_j - u_j v_i)$~\cite{Cooper99,Cooper01}. Consequently, a bosonic FQH state is distinguished from its fermionic counterpart by a different filling factor: $\nu_b^{-1}=\nu_f^{-1}-1$. Small droplets of $\nu=1/2$ bosonic Laughlin states have been realized in recent experiments on  ultracold atoms~\cite{Leonard2023} and photonic circuits~\cite{Wang2024}. 

\begin{figure}[tb]
    \centering
    \includegraphics[width=0.48\textwidth]{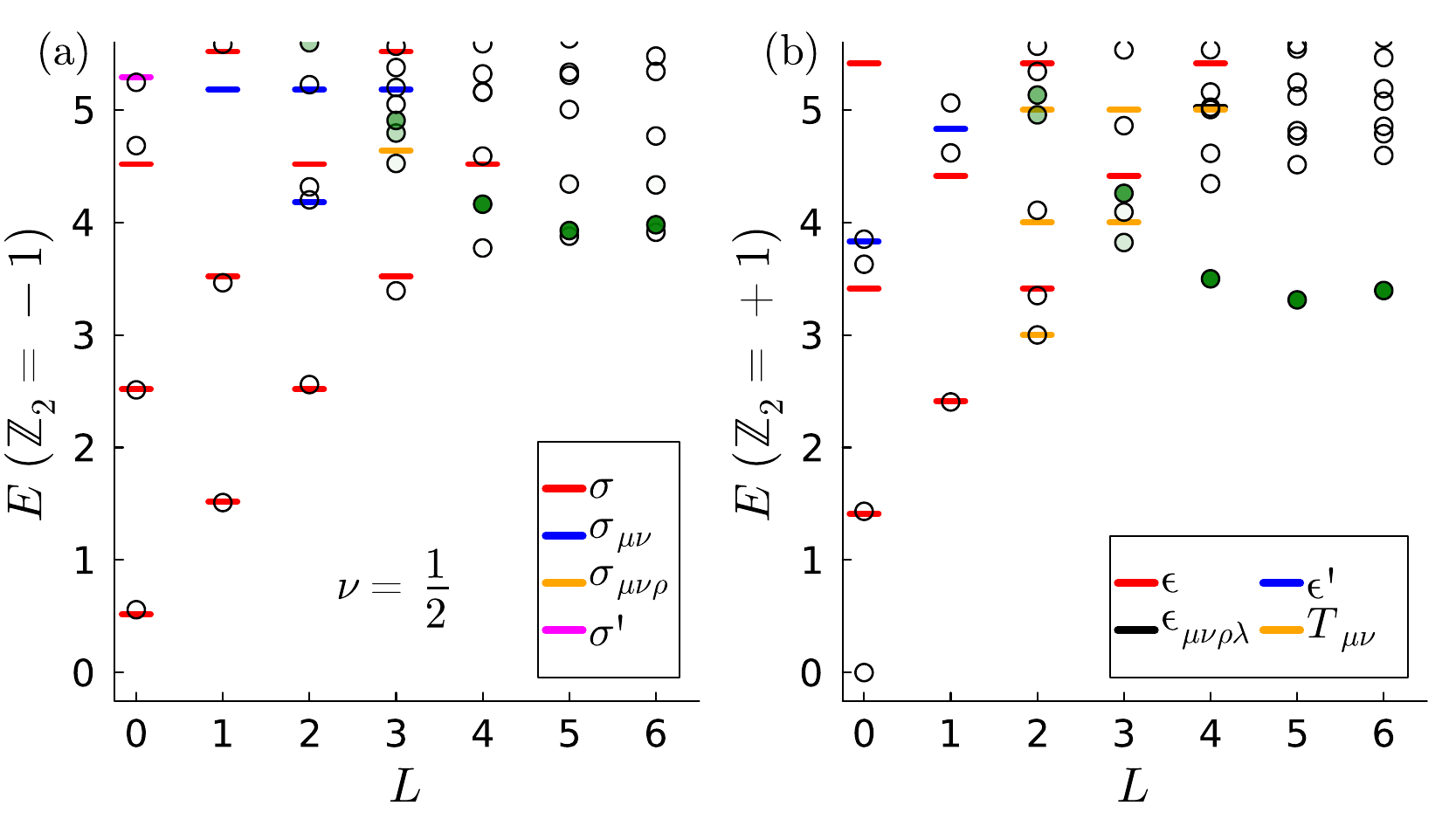}
    \includegraphics[width=0.48\textwidth]{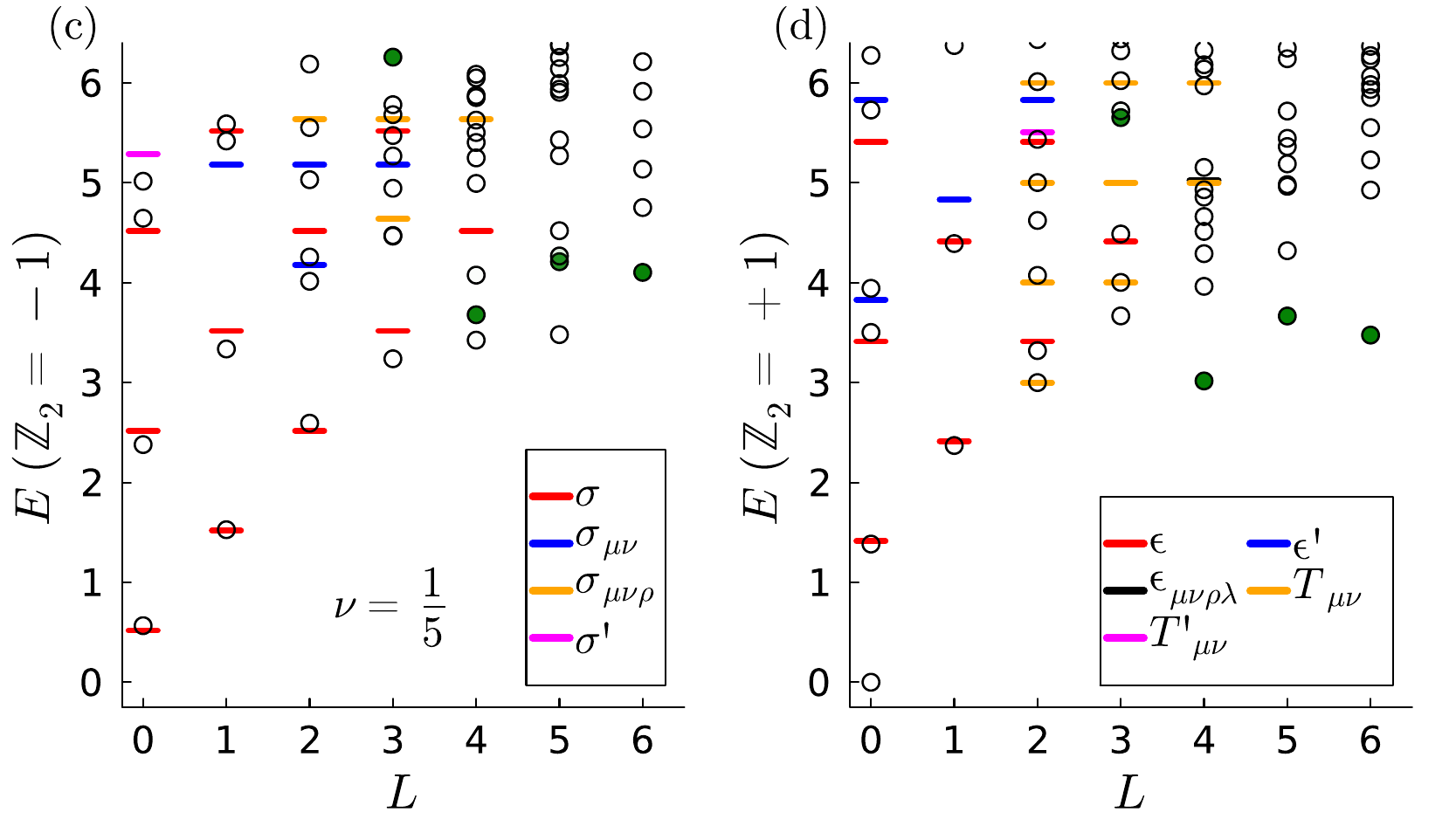}
    \caption{
    Spectra of the optimal models for other examples of Laughlin states. 
    	(a)-(b): Bosonic $\nu=1/2$ Laughlin state with $N=8$ particles at the optimal interaction point (see text). 
       (c)-(d): Fermionic $\nu=1/5$ Laughlin with $N=6$ electrons at the optimal interaction point (see text).
       	In all cases, the ED data are represented by empty circles. The states with high overlap with the MF magnetoroton are filled green. 
    	The spectra are in good correspondence with the conformal bootstrap data of the 3D Ising transition, indicated by line markers.  
    }\label{fig:conformal_spectrum_othernu}
\end{figure}

\subsection{Other models}\label{app:generalmodel}

We propose the following approach for finding models at fractional filling for which the 3D Ising transition can be observed. The starting point are single-layer Hamiltonians that realize the desired topological order, where $m$ is the highest relative angular momentum that is projected out. In the case of Laughlin states, for example, we have the relation $\nu = 1/(m+2)$. Let us fix the intralayer interaction according to
\begin{equation}\label{eq:appintra}
    V_p^\mathrm{intra} = \begin{cases}
        1 & p \leq m \\
        0 & \mathrm{otherwise,}
    \end{cases}
\end{equation}
while the fine tuning can be performed in the interlayer interaction. For simplicity, we preserve SU(2) layer symmetry up to the $m$th pseudopotential, and then add two more pseudopotentials, which are found through optimization. This ensures that, in the $h \to \infty$ limit, the ground state deviates as little as possible from a pure Laughlin paramagnet. Specifically, the interlayer interaction is
\begin{equation}\label{eq:appinter}
    V_p^\mathrm{inter} = \begin{cases}
        1 & p \leq m \\
        V_p & m < p \leq m+2 \\
        0 & \mathrm{otherwise.}
    \end{cases}
\end{equation}

Note that the original $\nu=1$ model from Ref.~\cite{Zhu23} also fits in this framework: since there is no energy scale set by the intralayer interaction, one can arbitrarily fix one of the inter-layer pseudopotentials, and optimize only over the remaining pseudopotential and the transverse field, i.e., two parameters in total. At any fractional filling, optimization needs to be done over three parameters instead. Below we showcase this method at fillings $\nu=1/2$ (bosons) and $\nu=1/5$ (fermions). We note that the optimal CFT points were empirically found to display similar pseudopotential ratios: $V_0/V_1 \approx 4.75$ ($\nu=1$), $V_1/V_2 \approx 5.88$ ($\nu=1/2$), $V_2/V_3 \approx 5.61$ ($\nu=1/3$) and $V_4/V_5 \approx 6.16$ ($\nu=1/5$), with a slight drift towards larger ratios as the filling factor is reduced. This suggests that the effective model in the spin space is similar in all cases, however at present we do not have a quantitative understanding of the relation between pseudopotential ratios and the filling factors, or why the proposed model works in general.

\begin{figure}[tb]
    \centering
    \includegraphics[width=0.48\textwidth]{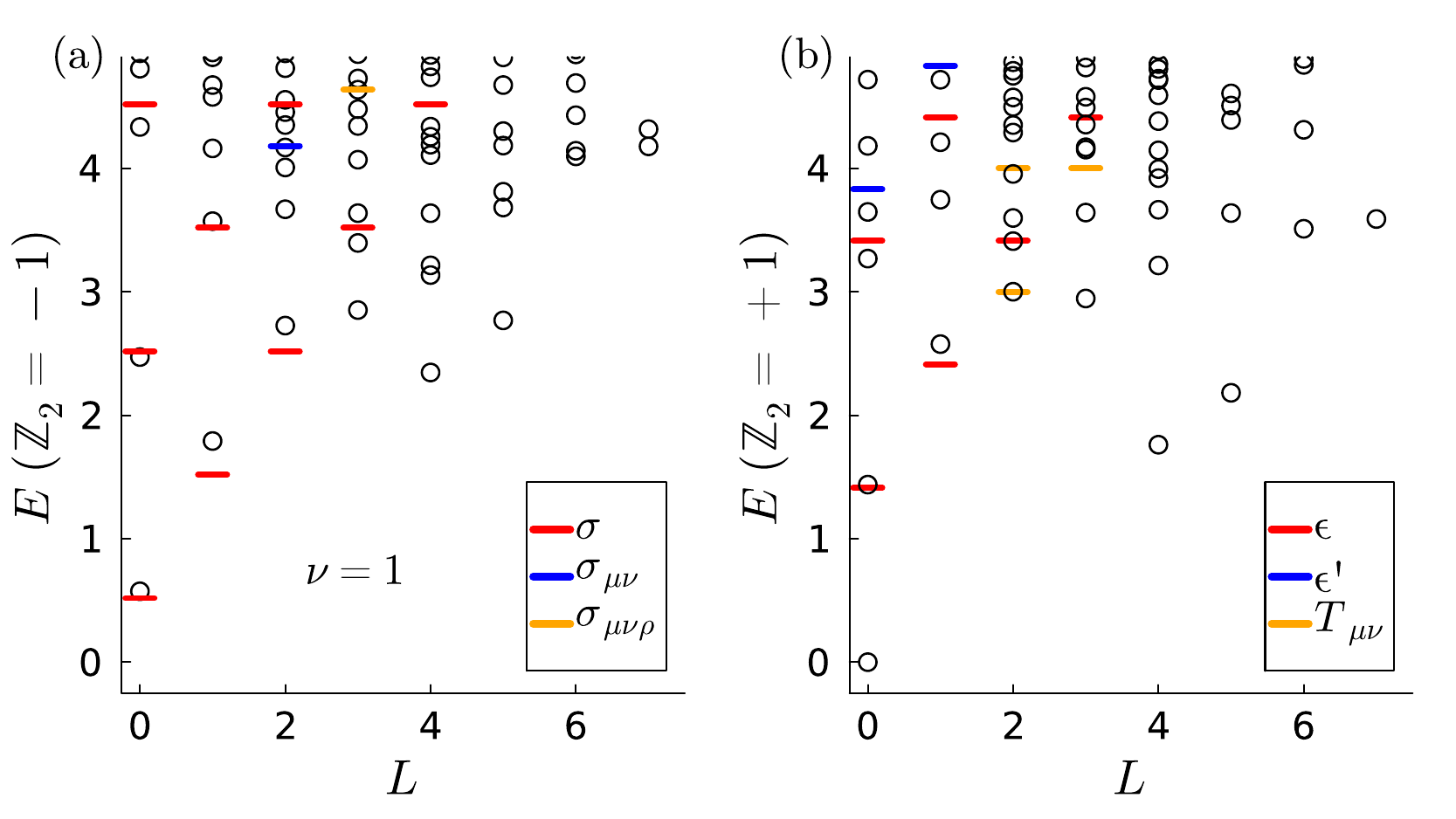}
    \includegraphics[width=0.48\textwidth]{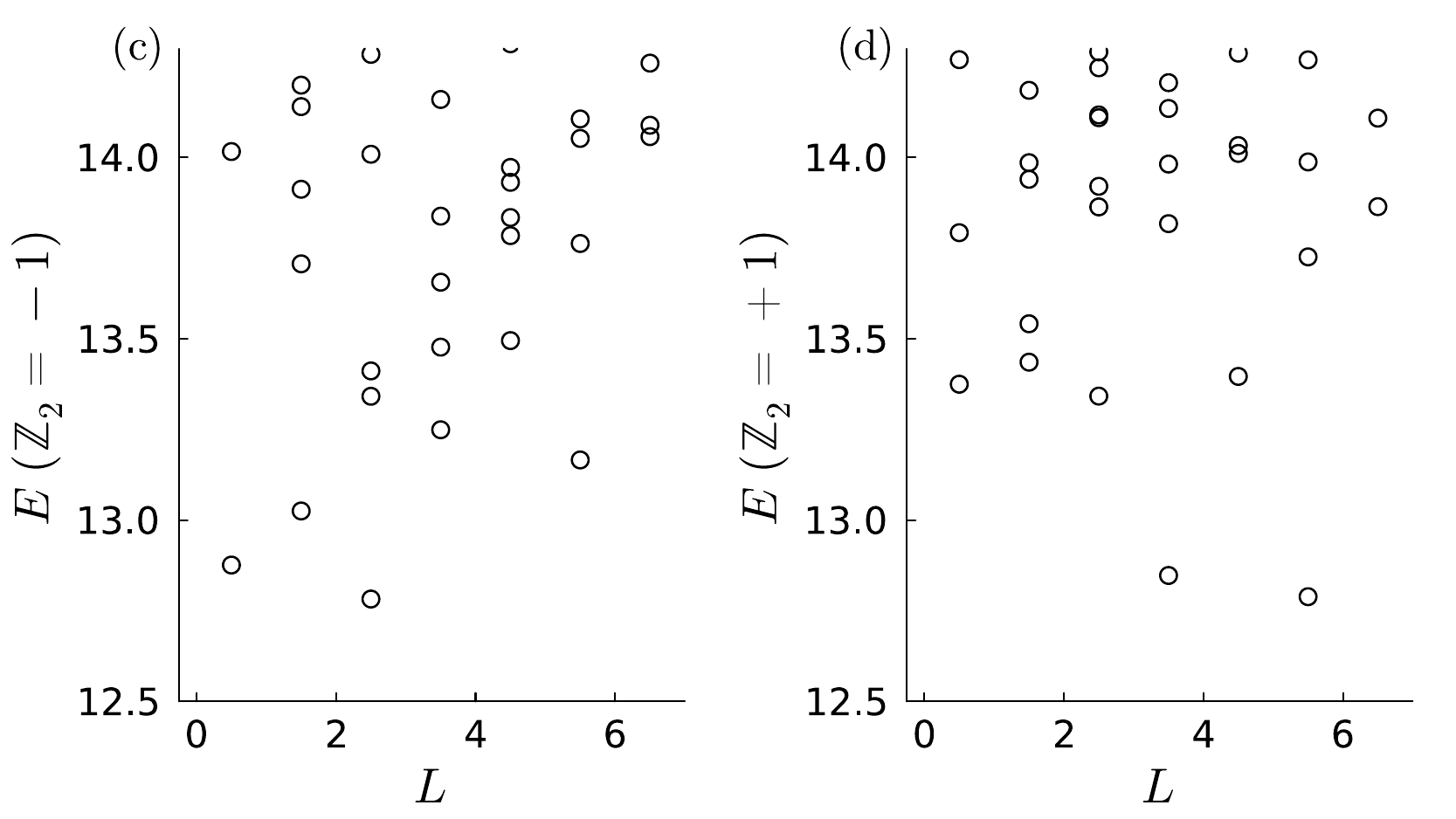}
    \caption{Spectra of the optimal model for the Moore-Read state of bosons at $\nu =1$. (a)-(b): Spectrum for $N=10$ bosons in $2Q=8$ flux quanta at the optimal interaction point (see text). The low-lying states of the spectrum match well against the conformal bootstrap data of the 3D Ising transition (shown in the legend). 
    (c)-(d): Spectrum for $N=11$ bosons in $2Q=9$ flux quanta. The angular momenta now take half-integer values. On the fuzzy sphere, the Moore-Read ground state is absent for odd numbers of particles and the CFT towers, similarly, could not be identified.
    }
    \label{fig:conformal_spectrum_pfaffian}
\end{figure}

\subsection{Some examples}\label{app:examples}

Here we provide some illustrations that our model, defined by~\cref{eq:appintra}-\cref{eq:appinter} above, works.  Using the same cost function as discussed in detail in \cref{app:optimization}, we optimize the spectrum for $N=8$ bosons at $\nu=1/2$ and find $V_1 = 0.49$, $V_2=0.10$, $h= 0.25$. \cref{fig:conformal_spectrum_othernu}(a)-(b) shows the spectrum at this special point, demonstrating good agreement with the 3D Ising based on the corresponding bootstrap data. Similarly, we optimized the spectrum for $N=6$ electrons at $\nu=1/5$ and found $V_4 = 0.37$, $V_5=0.06$, $h= 0.11$. \cref{fig:conformal_spectrum_othernu}(c)-(d) shows the spectrum at this special point, once again demonstrating good agreement with the bootstrap data. According to the optimized cost function, \cref{eq:cost_fn}, the finite-size effects at the critical point do not appear to significantly increase as the filling factor is decreased.

Finally, we demonstrate a non-Abelian model with bosons at filling $\nu=1$ featuring the 3D Ising CFT. It is known that, for spinless bosons at $\nu=1$, $V_0$ interaction can stabilize the Moore-Read Pfaffian state~\cite{Moore91,Regnault04}, whose quasihole excitations behave as Ising anyons with non-Abelian braiding statistics~\cite{Nayak96}.  We find that this state is also realized in our generic model above. Upon optimizing for the lowest-lying states of the CFT tower, we find the optimal parameters of $V_1 = 0.45$, $V_2 = 0.09$ and $h = 0.43$. In the limits of small and large $h$ field, the bosons form the Moore-Read state, as confirmed by the characteristic counting in the entanglement spectrum, corresponding to a chiral boson coupled to a Majorana fermion. Figures~\ref{fig:conformal_spectrum_pfaffian}(a)-(b) show the optimal spectrum in the even particle sector. While finite-size effects are stronger than for the previously studied Laughlin states, the low-lying CFT states for $L\leq 2$ can still be distinguished clearly.

One new feature in the non-Abelian case is the sensitivity to the parity of the particle number. Figures~\ref{fig:conformal_spectrum_pfaffian}(c)-(d) show the spectrum in the \emph{odd} particle number sector, for $N=11$ bosons and $2Q = 9$. The Moore-Read phase can be viewed as a $p+ip$ paired superconductor~\cite{Read00}, hence it does not have a ground state for odd particle numbers on the fuzzy sphere. Instead, its low-lying spectrum consists of a ``neutral fermion'' excitation~\cite{Bonderson11, Moller11}, which is interpreted as a Bogoliubov quasiparticle of the underlying $p+ip$ superconductor. Intriguingly, while the neutral fermion mode is visible in \cref{fig:conformal_spectrum_pfaffian}(c)-(d), there is no visible trace of the evenly-spaced CFT tower structure that is present in the even particle sector. We attribute this to the absence of a conformal vacuum for odd $N$, and it remains to be understood if any CFT information can be extracted in such cases.

\bibliography{bibliography}

\end{document}